\documentclass[usegraphicx,useAMS,usenatbib,usefloat]{mn2e}
\usepackage{txfonts}
\usepackage{graphicx}
\usepackage{epsfig}
\usepackage{longtable}
\usepackage{float}

\begin{document}
\title{Planetary Nebula Velocities in the Disk and Bulge of M31}
\author[Halliday et al.]
       {C. Halliday$^{1,2,3,4}$, D. Carter$^1$, T.J. Bridges$^{5,6}$, Z.C. Jackson$^1$,
       M.I. Wilkinson$^7$, 
\newauthor D.P. Quinn $^7$, N.W. Evans$^7$, N.G. Douglas$^8$, H.R. Merrett$^9$, M.R. Merrifield$^9$,
\newauthor  A.J. Romanowsky$^{10}$, K. Kuijken$^{8,11}$, M.J. Irwin$^7$.
       \\$^1$ Astrophysics Research Institute,
       Liverpool John Moores University, Twelve Quays House, Egerton
       Wharf, Birkenhead CH41 1LD, UK.
       \\$^2$ INAF - Osservatorio Astronomico di Padova, 
       Vicolo dell'Osservatorio 5, I-35122 Padova, Italy.
       \\$^3$ Institut f{\"u}r Astrophysik, Friedrich-Hund-Platz 1, 37077
              G\"{o}ttingen, Germany.
       \\$^4$ INAF - Osservatorio Astrofisico di Arcetri, Largo E. Fermi 5, 
              I-50125, Firenze, Italy.
       \\$^5$ Anglo-Australian Observatory, PO Box 296, Epping NSW 1710, Australia.
       \\$^6$ Department of Physics, Queen's University, Kingston, Ontario, K7L 3N6, Canada.
       \\$^7$ Institute of Astronomy, Madingley Road, Cambridge CB3 OHA, UK.
       \\$^8$ Kapteyn Astronomical Institute, Postbus 800, 9700 AV Groningen, The Netherlands.
       \\$^9$ School of Physics and Astronomy, University of Nottingham, University Park, 
        Nottingham, NG7 2RD, UK.
       \\$^{10}$ Departamento de F\'{i}sica, Universidad de Concepci\'{o}n, Casilla 160-C, 
       Concepci\'{o}n, Chile.
       \\$^{11}$ Leiden Observatory, P.O. Box 9513, NL-2300 RA  Leiden, The Netherlands.}

\date{Accepted ... .
      Received ... ;
      in original form ... }
\pagerange{\pageref{firstpage}--\pageref{lastpage}}
\pubyear{2006} \volume{000} \pagerange{1} \onecolumn

\maketitle \label{firstpage}

\begin{abstract}

We present radial velocities for a sample of 723 planetary nebulae
(PNe) in the disk and bulge of M31, measured using the
WYFFOS fibre spectrograph on the William Herschel telescope.
Velocities are determined using the [OIII]$\lambda$5007 emission line. 
Rotation and velocity dispersion are measured to a radius
of 50 arcminutes (11.5 kpc), the first stellar rotation
curve and velocity dispersion profile for M31 to such a radius. 
Our kinematics are consistent with rotational support 
at radii well beyond the bulge effective radius of 1.4kpc, although 
our data beyond a radius of 5kpc are limited. We present tentative
evidence for kinematic substructure in the bulge of M31 to be
studied fully in a later work. This paper is part of an ongoing
project to constrain the total mass, mass distribution and velocity
anisotropy of the disk, bulge and halo of M31.

\end{abstract}


\section{Introduction}

The Andromeda Galaxy M31 is our closest giant galaxy of any type,
and presents the best opportunity for conducting a robust analysis 
of the dynamics of a galaxy disk, bulge and halo. Early studies of the kinematics
of this galaxy were based upon the HI 21cm line velocities as a 
measure of rotation (Kent, 1989; Roberts \& Whitehurst, 1975) and HII 
emission line regions (Rubin \&
Ford, 1970). Using a combination of the rotation curve measures, and 
kinematics of the bulge component as measured from radial velocities
of 149 globular clusters, and assuming velocity isotropy, Kent {\sl et
al.} (1989) constrained the relative mass fractions of the disk and
bulge of M31.

The extremely rapid rotation seen in integrated stellar light velocity 
measurements in the centre 
(Kormendy 1988; Carter \& Jenkins 1993; van der Marel {\sl et al.} 1994;
Kormendy \& Bender 1999) and the double nucleus seen in HST images 
(Lauer {\sl et al.} 1993) have been interpreted as evidence for a central 
black hole of $1 - 3 \times 10^8 M_{\odot}$ (Bender {\sl et al.} 2005).
Models with a single rapidly rotating disk (Kormendy \& Bender 1999) or 
two nested disks (Bender {\sl et al.} 2005) have been proposed. The nucleus
of M31 is clearly complex.

Due to the steep decline in radial 
surface brightness profiles of the galaxy integrated light, alternative probes 
of the galaxy mass
distribution must be used at radii much greater than the bulge effective radius 
of 1.4 kpc (Pritchet \& van den Bergh 1994; Irwin {\sl et al.} 2005). The 
most recent HI 21cm observations
of M31 are summarised by Braun (1991), who concludes that the rotation curve 
is well fitted by a mass distribution which follows the stellar light 
with a fixed M/L of 6.5 solar in the B band. He finds no need for a massive dark halo to
a radius of at least 28 kpc. Nevertheless, spiral galaxies are widely believed 
to be dark-matter dominated in their outer parts, both to understand the 
majority of HI rotation curves, and to stabilise the disks (e.g. Ostriker, 
Peebles \& Yahil 1974).

Ideal probes of the mass distribution should trace the kinematics of the
galaxy disk, bulge and halo. With sufficiently large 
samples of kinematic measurements of these, it is possible to reconstruct the 
potential of a galaxy, and the distribution function of the tracer population 
for a spherical system (Merritt \& Saha 1993; Merritt 1993), or an axisymmetric 
two-integral galaxy viewed edge-on (Merritt 1996). Whilst neither of these 
approximations applies exactly to M31 (it is neither spherical nor edge-on), 
it is still possible to use discrete velocities to constrain self-consistent 
multi-component models (e.g. Widrow \& Dubinski 2005). 

Knowledge of the total mass of M31 is vital for the interpretation of the latest
generation of pixel microlensing experiments, which concentrate on that
galaxy (Crotts 1992; Kerins {\sl et al.} 2001, 2003; 
de Jong {\sl et al} 2004; Cseresnjes {\sl et al.} 2005;
Riffeser {\sl et al.} 2003; Belokurov {\sl et al.} 2005).  
Evans \& Wilkinson (2000) have found that there is no kinematic evidence to
support the widely-held belief that M31 is more massive than the Milky Way, 
in contradiction to previous studies. Their measurement is uncertain by a factor of
$\sim$2, the main source of uncertainty being the small number of
available tracer velocities particularly at large radii. Evans {\sl et al.}
(2000) presented radial velocity measurements for the 15 dwarf companions known at present,
and derived a total mass of M31 in the range $7 - 10 \times 10^{11} 
M_{\odot}$. 

Globular clusters have 
the advantage of being found at large galactocentric radii, but they are 
not present in large numbers in spiral galaxies.
Perrett {\sl et al.} (2002) presented radial velocities for 202 globular clusters,
mostly along the disk of M31. Kent {\sl et al.} (1989) and Huchra {\sl et al.} 
(1991) presented 149 globular cluster radial velocities, including some from older
sources. Smaller samples are presented by Federici {\sl et al.} (1993; 35 globular clusters), 
and Jablonka {\sl et al.} (1998; 16 clusters). With some duplication, the total 
sample of globular cluster velocities is now of order 300. Evans {\sl et al.}
(2003) combine the globular cluster velocities from Perrett {\sl et al.} with
the dwarf satellite velocities, and deduce that M31 has a dark halo which is
isothermal out to at least 100 kpc, and has a total mass within that radius
of $ \sim 1.2 \times 10^{12} M_{\odot}$. 

Planetary nebulae (PNe) are not as spatially extended as globular 
clusters. They are readily detectable however due to their bright [OIII] emission lines, and are 
substantially more numerous than globular clusters. Their spatial 
distribution in M31 is expected to reflect that of the old stellar population of
the disk and bulge. 
Pritchet \& van den Bergh (1994) find that the distribution of the bulge stellar population
can be fit by a single 
de Vaucouleurs (1948) law from the inner bulge at a radius of 200 parsecs, out 
to $\approx$20 kpc, beyond which the profile flattens to an exponential or a power 
law of index about -2.3 (Irwin {\sl et al.} 2005). 
PNe are used as tracers of galaxy kinematics using two different methods: PNe are 
identified in narrow band, wide field [OIII] images, and then their velocities 
measured subsequently with a multi-object spectrograph (e.g Arnaboldi
{\sl et al.} 1996; Arnaboldi {\sl et al.} 1998; Hui {\sl et al.} 1995;
Peng {\sl et al.} 2004); or a slitless spectrograph combined 
with narrow-band filters is used to complete imaging and spectroscopy in a single
observation, albeit with a narrower field of view (Douglas {\sl et al.} 
2000, 2002; M\'{e}ndez {\sl et al.} 2001). 

Hurley-Keller {\sl et al.} (2004, hereafter HK04) present radial velocities 
of a sample of 135 PNe to the South and East of the nucleus of M31. They find that
most of the PNe in their sample belong to rotationally supported disk and bulge 
components, with no evidence for a dynamically hot halo, although the sample of
PNe they have studied is rather small.

Here we present accurate velocities for 723 PNe observed using the 
first of these two approaches, using the WYFFOS multi-object spectrograph on
the William Herschel Telescope on La Palma. In section \ref{astrometry}
we describe methodology for object selection and astrometry; in section
\ref{spectroscopy} we describe the spectroscopic observations and basic data
reduction; in section \ref{velocities} we outline our measurement of PNe radial velocities and
determination of velocity errors. In section \ref{analysis} we present a kinematic
analysis of the disk and bulge of M31.
A survey of a larger number of PNe has been 
completed using the Planetary Nebula Spectrograph (Douglas {\sl et al.}
2002). Using a subsample of these velocities, Merrett {et al.} (2003)
trace the kinematics of the southern stellar stream of M31 through
the body of the galaxy, and suggest a possible stream orbit.
The complete velocity dataset from the Planetary Nebula Spectrograph,
together with a detailed comparison with the current dataset 
will be presented by Merrett {\sl et al.} (2006). A  further
paper will present a detailed astrophysical analysis of the joint dataset. 

\section{Observations and Data Reduction}

\subsection{Sample and Astrometry}
\label{astrometry}

Our spectroscopic data were acquired during two observing runs at the William
Herschel Telescope, in August/September 1999 and in October 2001.
Between our spectroscopic runs we completed an [OIII]+ Str\"{o}mgren y
imaging survey using the Isaac Newon Telescope to detect additional PNe
targets out to a projected radius of 20kpc. Here and throughout this paper
we assume a distance of 770 kpc for M31.
The sample selection and astrometry
thus differed between the 1999 and 2001 runs, and we describe each in turn.

The 1999 sample is derived from an original sample of 429 PNe detected by
Ciardullo {\sl et al.} (1989) (hereafter C89) from [OIII] and
continuum images of M31 taken in an irregularly shaped region
approximately 33 x 3 arcminutes in the disk of M31. C89 quote positions for equinox and epoch
B1975 and the internal precision of the positions is good. In a
successful multi-fibre observing run the telescope is aligned on the
target field using bright fiducial stars since the PNe are not visible
to an acquisition camera. The purpose of our astrometric procedure was
to ensure that our PNe targets and fiducial stars were on a common system, aligned 
with the International Celestial Reference Frame.

The raw material for our astrometric solutions were archive
images from the INT Wide Field Survey (WFS), particularly V band
images of two fields studied for the INT M31 pixel microlensing
project. Each exposure was 599 seconds in V for the four separate CCDs
of the INT Wide Field Camera (WFC).

It was first necessary to set up a secondary astrometric reference
grid on the WFC CCD images. This was completed using Digital Sky
Survey\footnote{The Digitized Sky Survey was produced at the Space
Telescope Science Institute under U.S. Government grant NAG
W-2166. The images of these surveys are based on photographic data
obtained using the Oschin Schmidt Telescope on Palomar Mountain and
the UK Schmidt Telescope. The plates were processed into the present
compressed digital form with the permission of these institutions.}
(DSS) images. A sample of 42 primary astrometric standards was chosen
from the PPM catalogue (Monet {\sl et al.} 1996). Using catalogued positions and
positions measured on the frame, an astrometric solution for the DSS
image was derived using the Starlink\footnote{The Starlink project is
run by CCLRC on behalf of PPARC}/AAO astrometric transformation
programme {\tt ASTROM}, written by P.T. Wallace. Positions of
secondary standards were then measured from the DSS images, and
transformed to the primary astrometric reference frame.

The astrometric solutions were derived from the secondary
standards. Between 35 and 40 secondary standards were measured on each
of the four CCD images in each of the two WFC fields of M31. Using
these standards, again using {\tt ASTROM}, and using a model of the
geometrical distortion at the prime focus of the INT provided by
R.W. Argyle, an astrometric solution was derived for each CCD image.

Because the WFC images were continuum the majority of the PNe were
not visible on them, but the brightest few were. The C89 positions are
derived from a set of reference stars given in Ford and Jacoby (1978)
(FJ78). To transform the C89 positions to our reference frame we
assumed that they were measured correctly with respect to the FJ78
standards, and measured the offset between the FJ78 positions
precessed to J2000, and our measured J2000 positions of the FJ78
stars. These offsets were of order 1 arcsecond in RA and Dec, with a
scatter of 0.4 arcsec. The consistency of the procedure was checked by
measuring 4 bright PNe in the C89 sample on the WFC frames.

Our catalogue for the 1999 run comprises the C89 catalogue 
precessed to J2000, and with the offset derived above applied, 
together with measured positions of fiducial stars suitable 
for use with WYFFOS.

For the 2001 sample both the PNe and fiducial star positions were 
determined from a set of [OIII] narrow band and Str\"{o}mgren y continuum images
obtained between August 3$^{rd}$ and 8$^{th}$ 2000 with the INT 
Wide Wield Camera (WFC). The Str\"{o}mgren y filter
defines a convenient continuum band close in wavelength to, but not
containing, the [OIII] lines. A list of 1284 potential targets was drawn 
from 12 sets of WFC images, covering a field of approximately 1.6 degrees
in Declination (Dec) by 1.4 degrees in Right Ascension (RA), centred on M31. Each 
set consisted of three 1200 second exposures in [OIII] and three 300
second exposures in Str\"{o}mgren y. After reduction some of these images 
were rejected because of problems with image quality, so the depth of
the images was not uniform from field to field. Remaining images were
bias subtracted, flatfielded and coadded, to provide one deep frame
in each passband. Corresponding coadded [OIII] and Stromgren y images were then 
compared visually using the blink
function of the Starlink {\tt GAIA} image display package. PNe were 
identified as objects appearing in the [OIII] frame but not the 
Str\"{o}mgren y frame, and their pixel co-ordinates were measured using 
the centroiding algorithm of the aperture photometry function of 
{\tt GAIA}. Positions of fiducial stars were measured from the [OIII]
frames in the same way.

Pixel co-ordinates were then transformed to ICRF J2000 RA and Dec using 
the same procedures as for our 1999
dataset. A set of secondary standards was set up on each WFC CCD image using
a DSS image, and the positions of the PNe and fiducials were 
calculated from these secondary standards, each time using {\tt ASTROM}. Our
final positions of PNe and fiducials have a relative precision of around 
0.4 arcseconds, although the absolute accuracy of our astrometric frame is 
around 1 arcseconds. Merrett {\sl et al.} (2006) present a comparison of the 
astrometry from this paper with that derived independently from PN.S data. They find 
that the combined error $\sigma = (\sigma_{PNS}^2 + \sigma_{H06}^2)^{1/2}$ = 0.48 arcsec
in RA, where H06 refers to the current paper,
with a lower dispersion in Dec but with a systematic offset of 1 arcsecond
between the positions in the central fields. This offset could be in either dataset. 

From the positions of the PNe which were successfully observed in both runs, 
we find that the 1999 and 2001 co-ordinate frames are consistent, the formal offset 
is that the 1999 positions are 0.03$\pm$0.02 seconds of time lower in RA, and 
0.1$\pm$0.1 arcseconds higher in Dec, than the 2001 positions. Considering the 
offset between the C89 positions, transformed to J2000, and the 2001 positions
we find that the offset is dependent upon position in the field. Over
that part of the field covered by the FJ78 secondary standards, the C89 positions
again are consistent with ours, formally the C89 positions are 0.02$\pm$0.01 seconds 
of time lower in RA, and 0.1$\pm$0.1 arcsec higher in Dec than our 2001 positions.
However C89 note that their positions are less reliable in the NE fields not covered by
FJ78 secondary standards, and in these fields we find that C89 positions are 0.13$\pm$0.04
seconds of time higher in RA, and 1.2$\pm$0.2 arcseconds higher in Dec than our 2001
positions (i.e. an offset of about 1.5 arcseconds on the sky).

In order to reduce contamination of the sample by HII regions, extended objects were
not selected. It is possible that some compact HII regions remain in the sample. Except
at the brightest magnitudes it is not possible to separate HII regions from PNe using
narrow band imaging alone (Merrett {\sl et al.} 2006), and this would only be 
possible using flux calibrated spectra with a wide wavelength range. 

In table \ref{tab:fiducials} we present the positions of a representative subset of 
the fiducial stars in the astrometric system of this paper. These positions could be used 
for fiducials for future fibre spectrograph runs, or as tertiary standards for astrometry
of further objects in the M31 field. Our full astrometric database of 1284 PNe positions, 
and 313 potential fiducial stars, is available in tabular form from either of the first 
two authors.

\begin{table}
\caption{\label{tab:fiducials}
Positions of a representative selection of fiducial or tertiary astrometric standard
stars}
\begin{center}
\begin{tabular}{lll} \hline
Identifier&RA&Dec \\
&(J2000)&(J2000) \\ \hline
FD\_12\_2\_9& 0 38 \ 1.350& 40 49 24.38\\
FD\_7\_2\_7& 0 38 \ 9.424& 41 51 \ 1.33\\
FD\_7\_4\_8& 0 38 36.612& 42 \ 2 34.67\\
FD\_8\_3\_6& 0 38 43.761& 41 34 27.69\\
FD\_7\_3\_4& 0 38 44.420& 42 \ 4 11.44\\
FD\_7\_1\_7& 0 38 45.846& 41 40 21.24\\
FD\_12\_3\_1& 0 38 47.375& 41 \ 5 \ 1.19\\
FD\_12\_3\_2& 0 39 \ 1.388& 41 \ 2 53.71\\
FD\_12\_4\_2& 0 39 11.011& 40 46 \ 8.65\\
FD\_12\_3\_4& 0 39 32.871& 41  1 54.33\\
FD\_12\_3\_6& 0 40 \ 4.205& 41  4 17.88\\
FD\_7\_3\_7& 0 40 \ 4.473& 42 13 32.58\\
FD\_11\_1\_2& 0 40 57.999& 41 43 52.33\\
FD\_11\_3\_8& 0 41 13.546& 42 \ 6 \ 3.83\\
FD\_6\_2\_3& 0 42 \ 2.068& 41 13 51.53\\
FD\_6\_2\_7& 0 42 \ 8.981& 41 23 30.29\\
FD\_6\_2\_2& 0 42 20.216& 41 \ 7 46.82\\
FD\_6\_2\_5& 0 42 30.763& 41 10 \ 9.66\\
FD\_11\_4\_7& 0 42 35.463& 41 57 46.87\\
FD\_10\_4\_8& 0 42 41.726& 41 55 33.18\\
FD\_6\_3\_1& 0 43 \ 1.609& 41 30 \ 7.62\\
FD\_6\_4\_3& 0 43 14.718& 41 24 59.03\\
FD\_6\_1\_2& 0 43 32.620& 41 \ 9 \ 9.31\\
FD\_10\_4\_3& 0 43 41.653& 41 53 11.48\\
FD\_5\_2\_10& 0 43 46.899& 41 26 11.35\\
FD\_10\_4\_1& 0 44 11.361& 42 \ 2 44.81\\
FD\_9\_2\_5& 0 44 22.363& 41 54 \ 0.15\\
FD\_4\_2\_2& 0 44 30.108& 40 46 29.34\\
FD\_6\_1\_3& 0 44 30.574& 41 15 37.17\\
FD\_6\_3\_3& 0 44 35.570& 41 30 36.09\\
FD\_4\_1\_1& 0 44 47.807& 40 34 22.34\\
FD\_9\_3\_1& 0 45 \ 8.474& 42 \ 9 58.52\\
FD\_9\_1\_8& 0 45 34.204& 41 41 43.99\\
FD\_4\_3\_6& 0 45 38.023& 40 57 38.22\\
FD\_5\_4\_8& 0 45 43.785& 41 19 43.98\\
FD\_9\_1\_7& 0 45 45.974& 41 49 20.30\\
FD\_4\_3\_2& 0 45 46.630& 40 56 47.39\\
FD\_4\_4\_10& 0 46 \ 4.986& 40 46 \ 4.43\\
FD\_4\_4\_9& 0 46 \ 9.054& 40 51 14.29\\
FD\_4\_4\_11& 0 46 36.072& 40 45 39.41\\ \hline
\end{tabular}
\end{center}
\end{table}

\subsection{Spectroscopic Data and Reduction}
\label{spectroscopy}

Spectra were obtained of PNe targets in two separate spectroscopic runs
with the WYFFOS fibre fed spectrograph (Worswick {\sl et al.} 1995)
on the William Herschel Telescope. WYFFOS was fed from the AutoFib 2
robotic fibre positioner (Parry {\sl et al.} 1994). Targets were 
selected to maximize the number of objects observed using the
configuration program {\tt AF2\_configure}. The minimum
object to object spacing of 25 arcseconds means that towards the centre of 
M31 many objects cannot be observed. There is however no lack
of targets in the central region.

Our wavelength range of
$\sim$350\AA~ was centred on the [OIII]$\lambda$4959\AA~and
[OIII]$\lambda$5007\AA~ emission lines, using the WYFFOS echelle
mode in 5th order. Four different 20 minute
exposures were made consecutively for each configuration. Flat-field
and neon comparison arc lamp exposures were obtained separately for
each configuration.

On the nights of 30$^{th}$ August to 2$^{nd}$ September 1999 we observed 12 different 
WYFFOS fibre configurations based upon the input catalogue from C89, 
yielding a total of 294 PNe spectra. During this run we used the 
original ``large'' fibre module of WYFFOS, with 2.7 arcsecond fibres,
and the consequent spectral resolution was 0.9 \AA. Between 24$^{th}$ and 28$^{th}$
October 2001 we acquired data for a further 12 WYFFOS configuration to a radius of 
11.5 kpc based on an input catalogue of 1284 PNe created from our INT WFC data. During 
this run we used the ``small'' fibre module of WYFFOS, with 1.6 arcsecond fibres,
and the consequent spectral resolution was 0.5 \AA.
These data provided a further 482 PNe spectra. The efficiency of this run at detecting 
faint PNe was however lower, owing to the smaller fibre sky area.  

\begin{figure*}
\begin{center}
\begin{minipage}{6in}
\epsfig{file=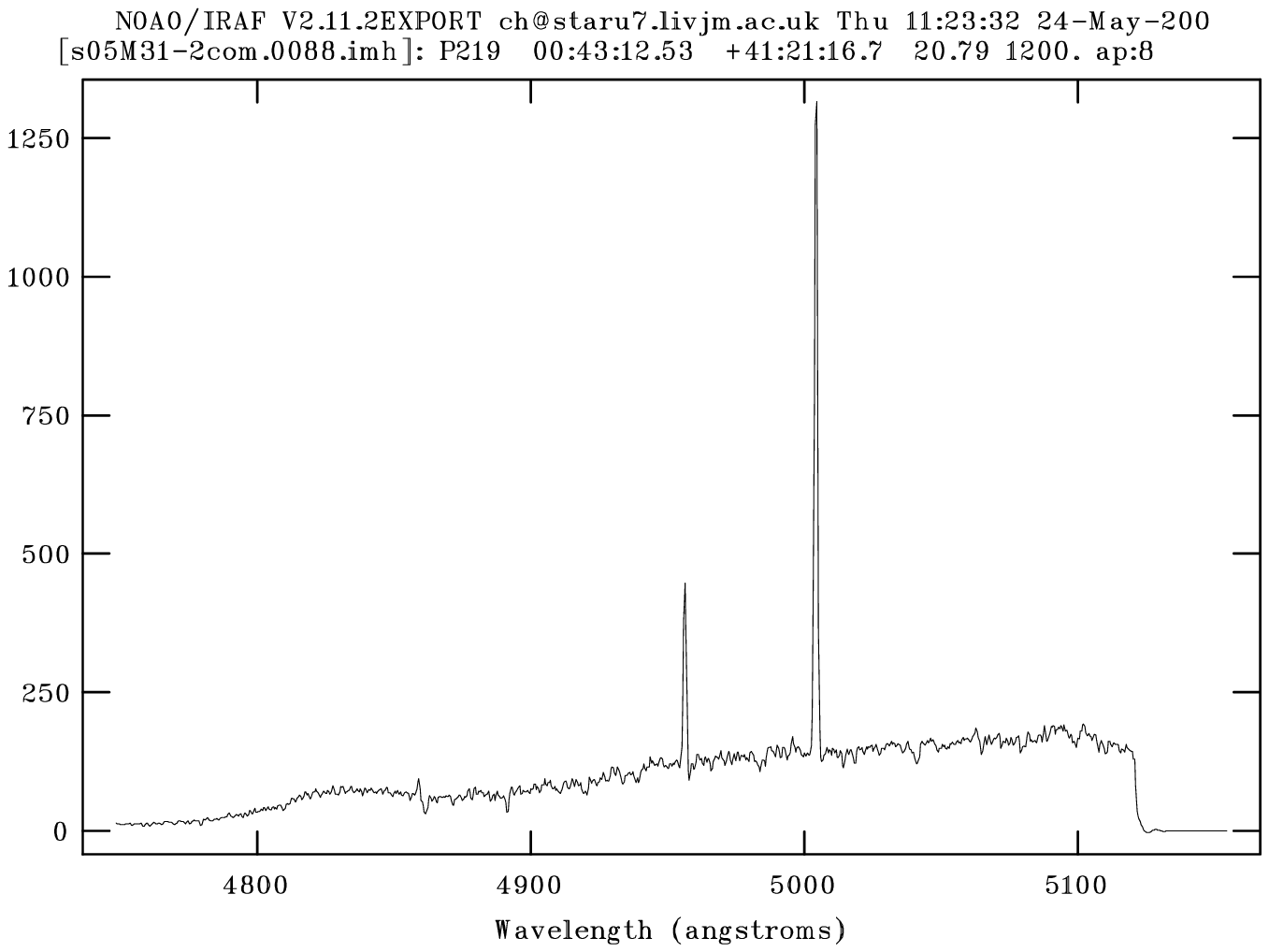,width=5.0cm,height=3.5cm}
\epsfig{file=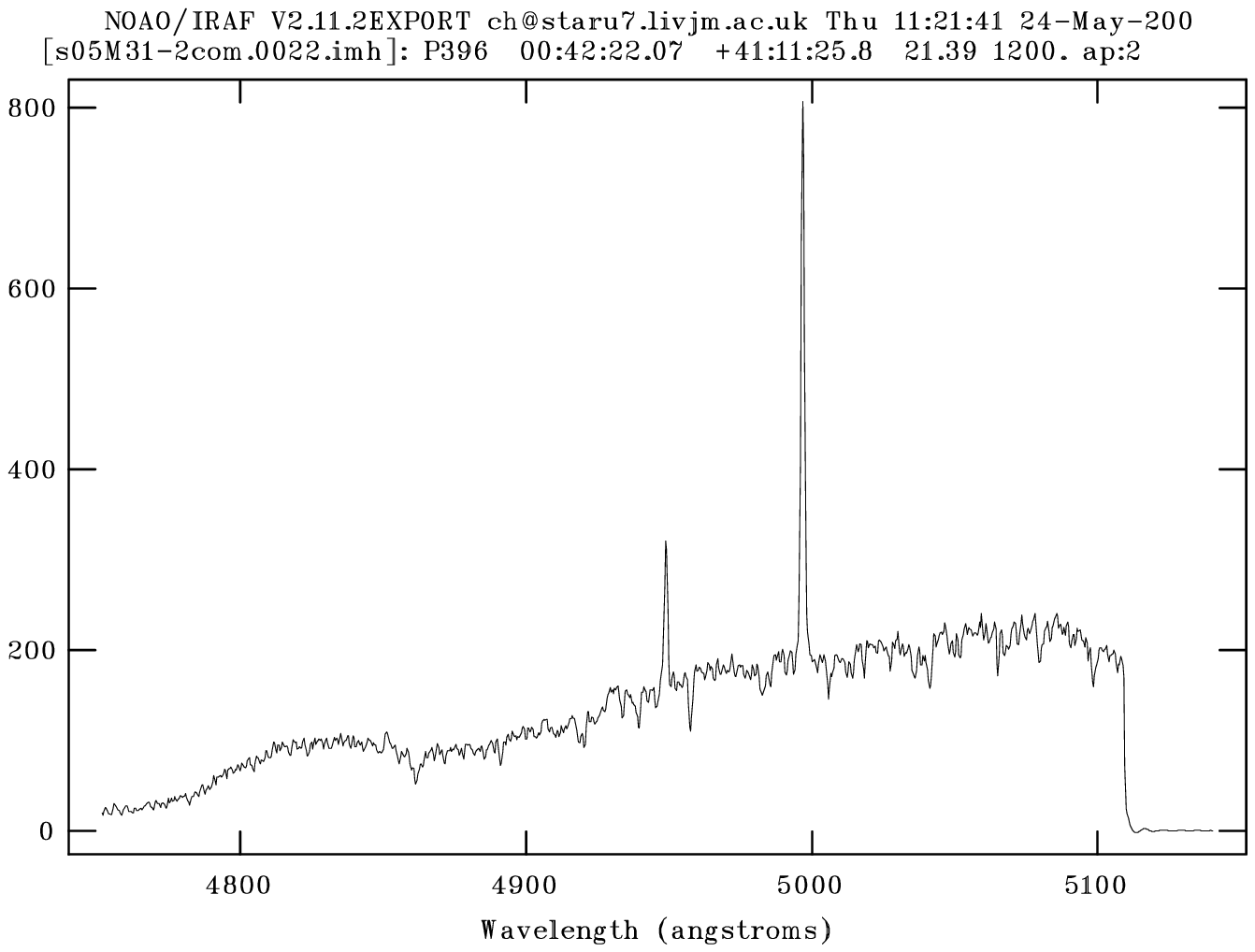,width=5.0cm,height=3.5cm}
\epsfig{file=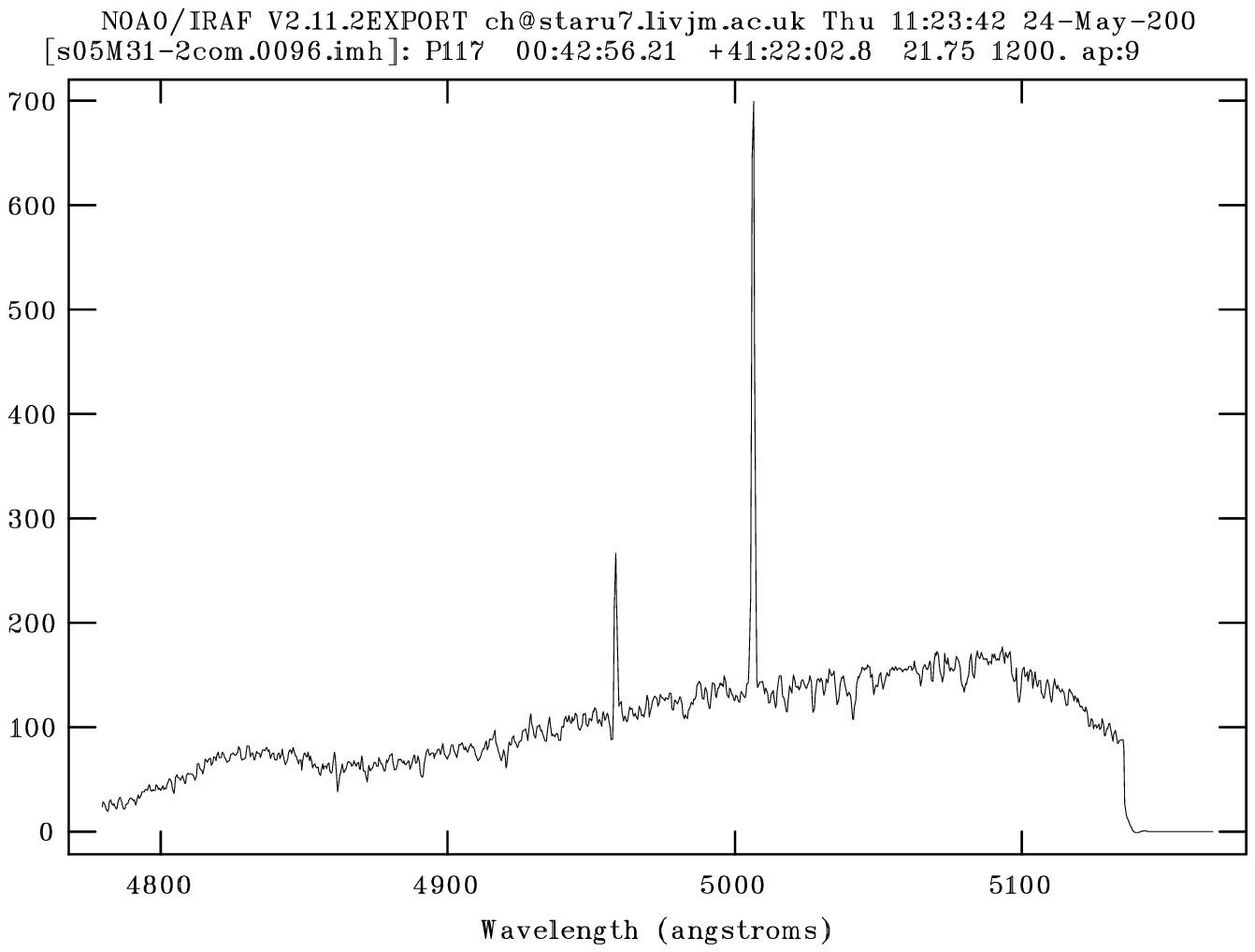,width=5.0cm,height=3.5cm}
\epsfig{file=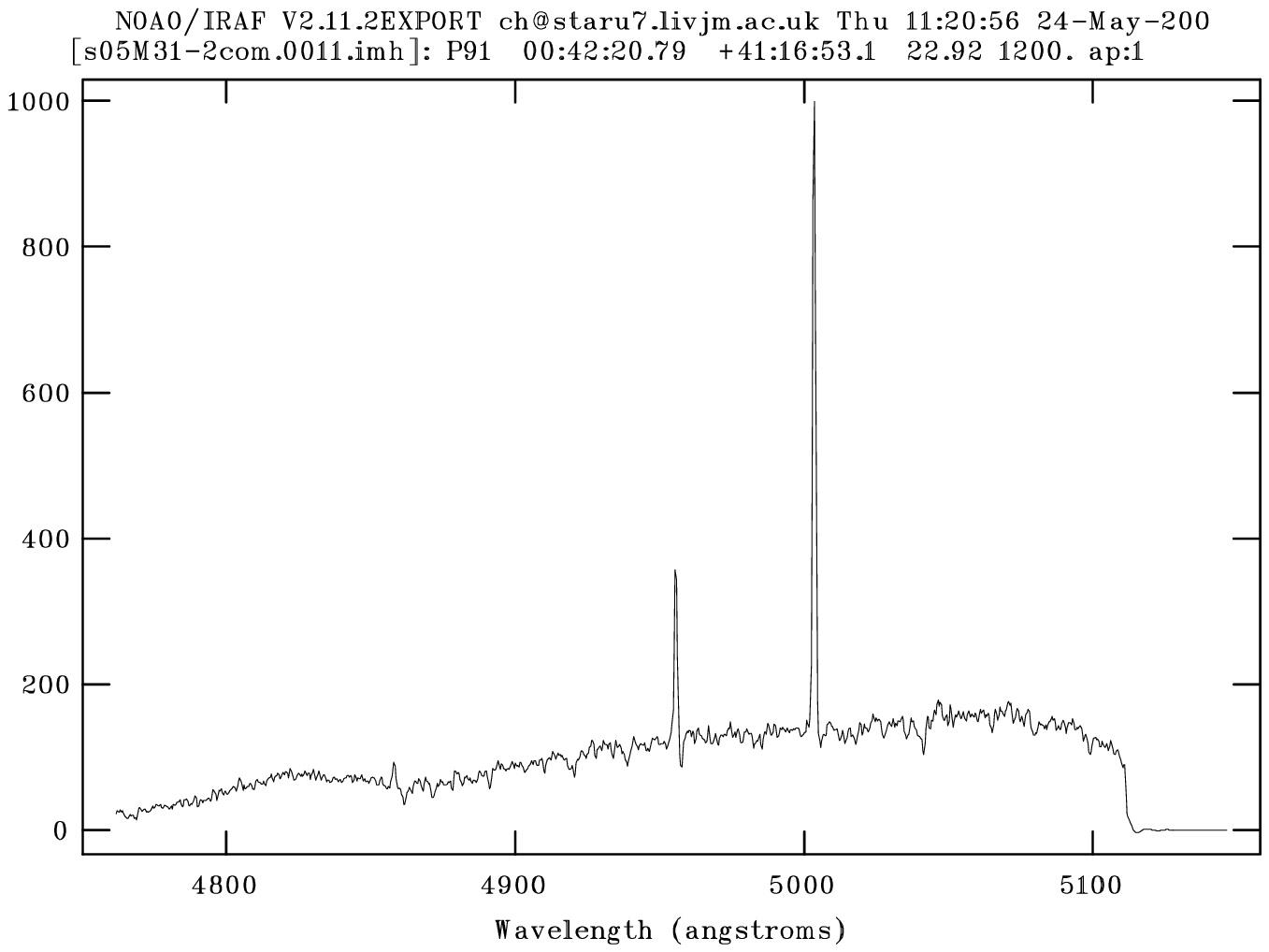,width=5.0cm,height=3.5cm}
\epsfig{file=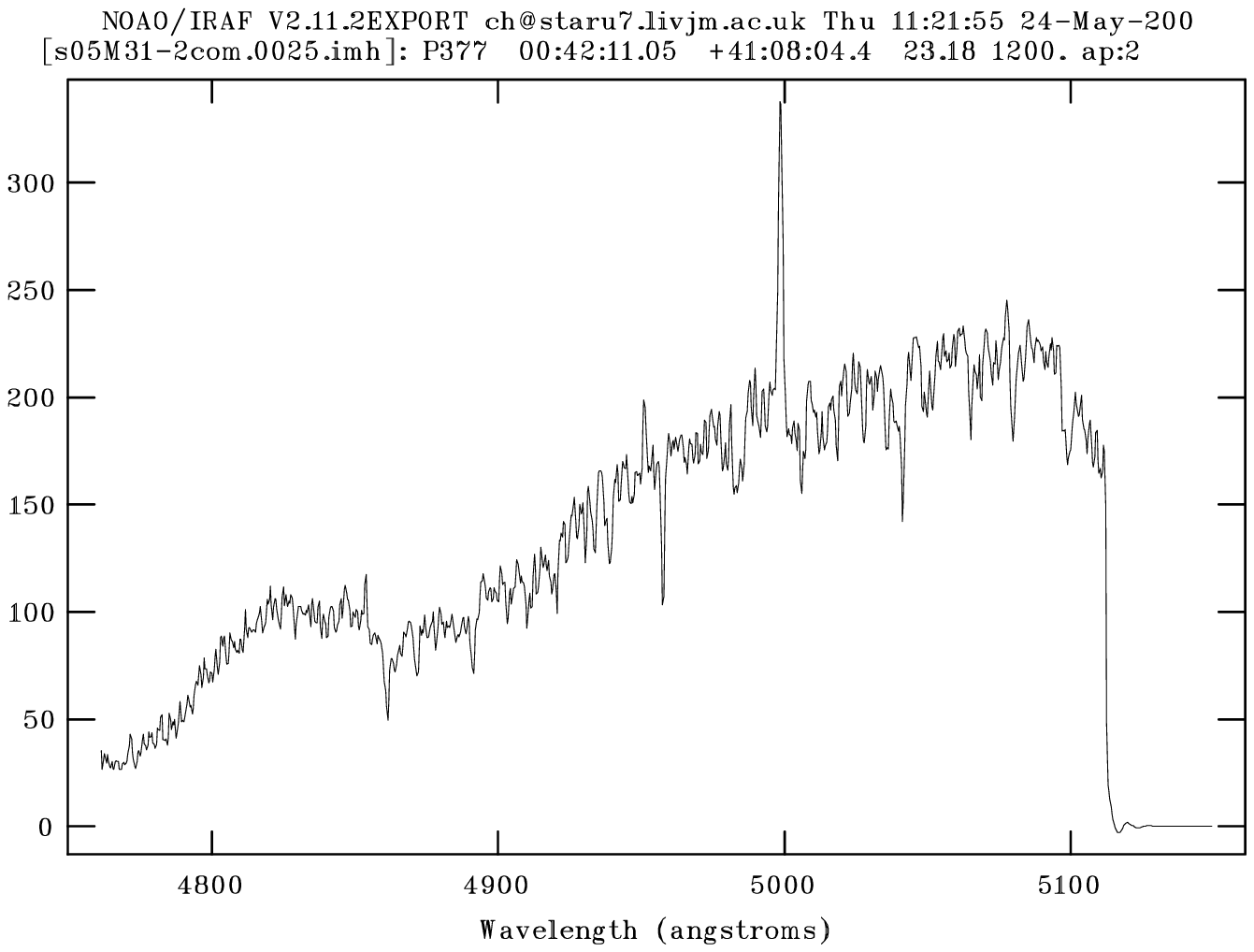,width=5.0cm,height=3.5cm}
\epsfig{file=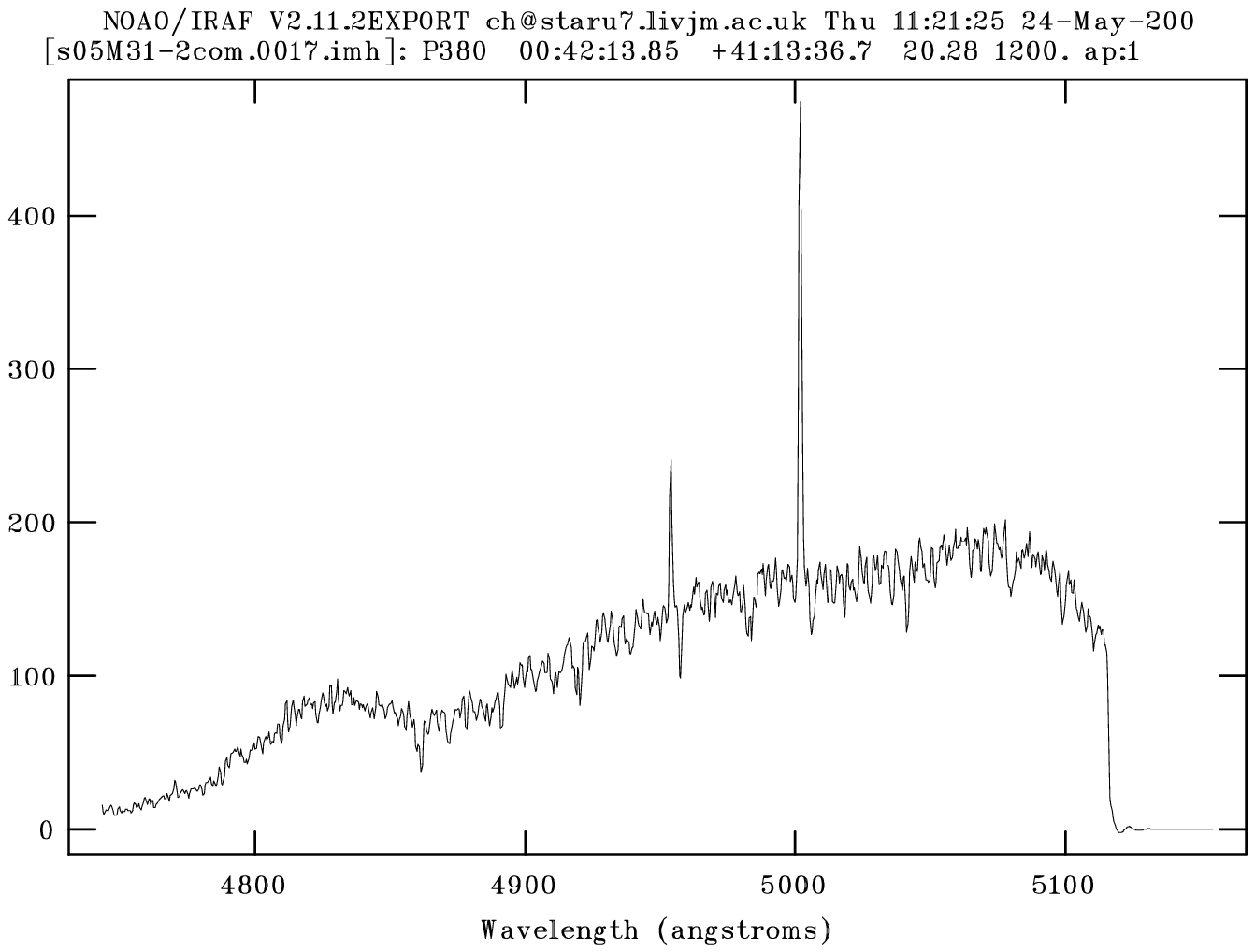,width=5.0cm,height=3.5cm}
\epsfig{file=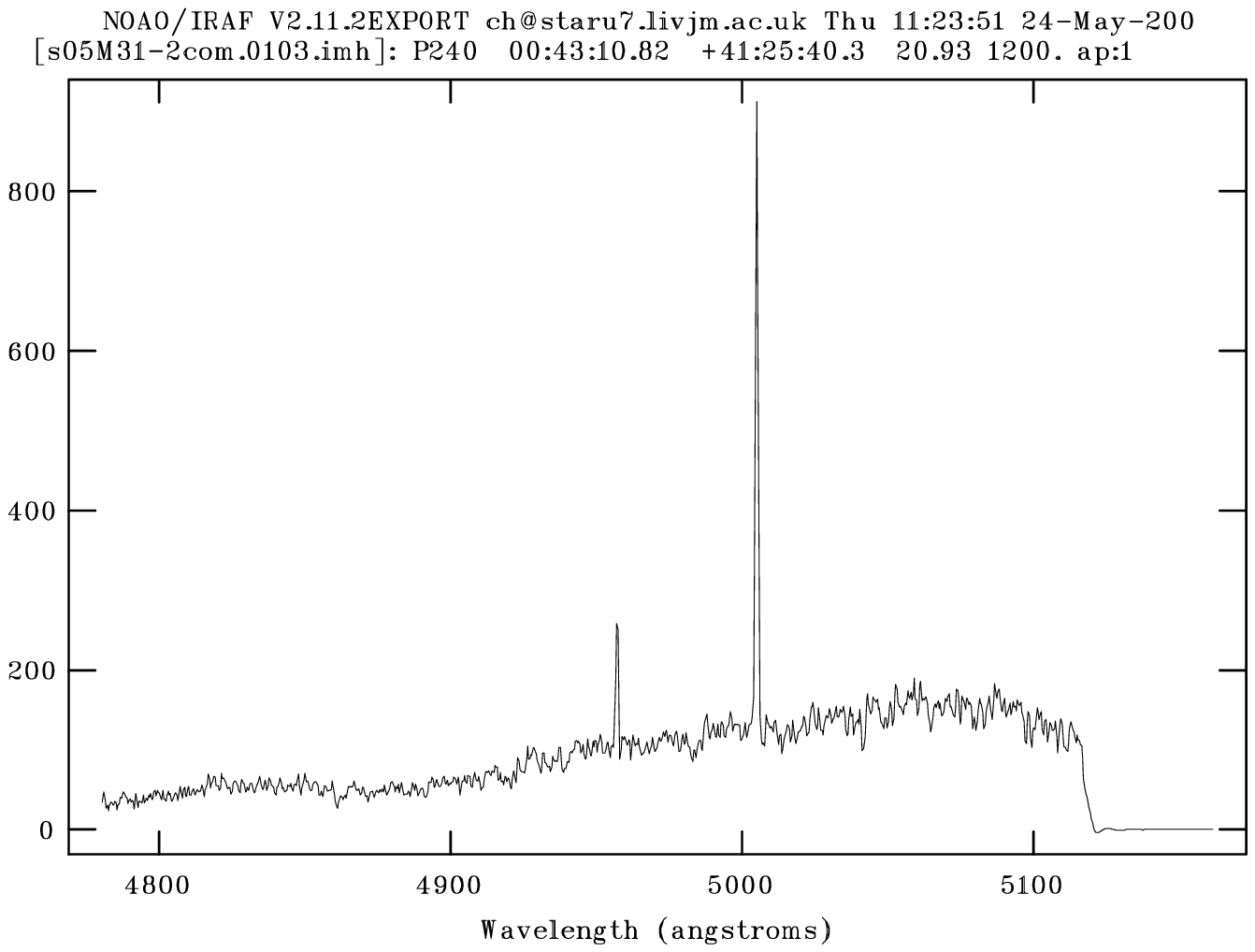,width=5.0cm,height=3.5cm}
\epsfig{file=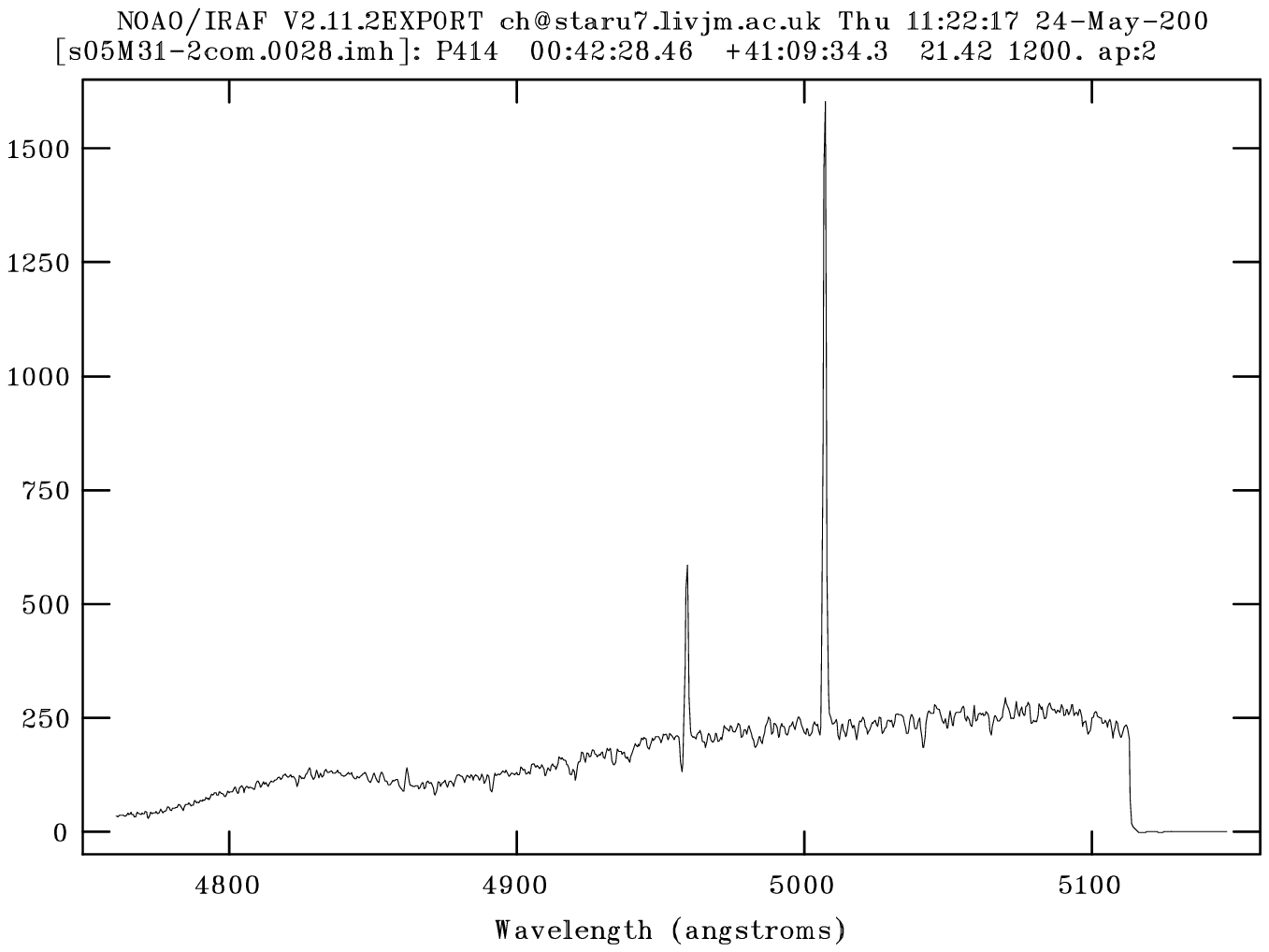,width=5.0cm,height=3.5cm}
\epsfig{file=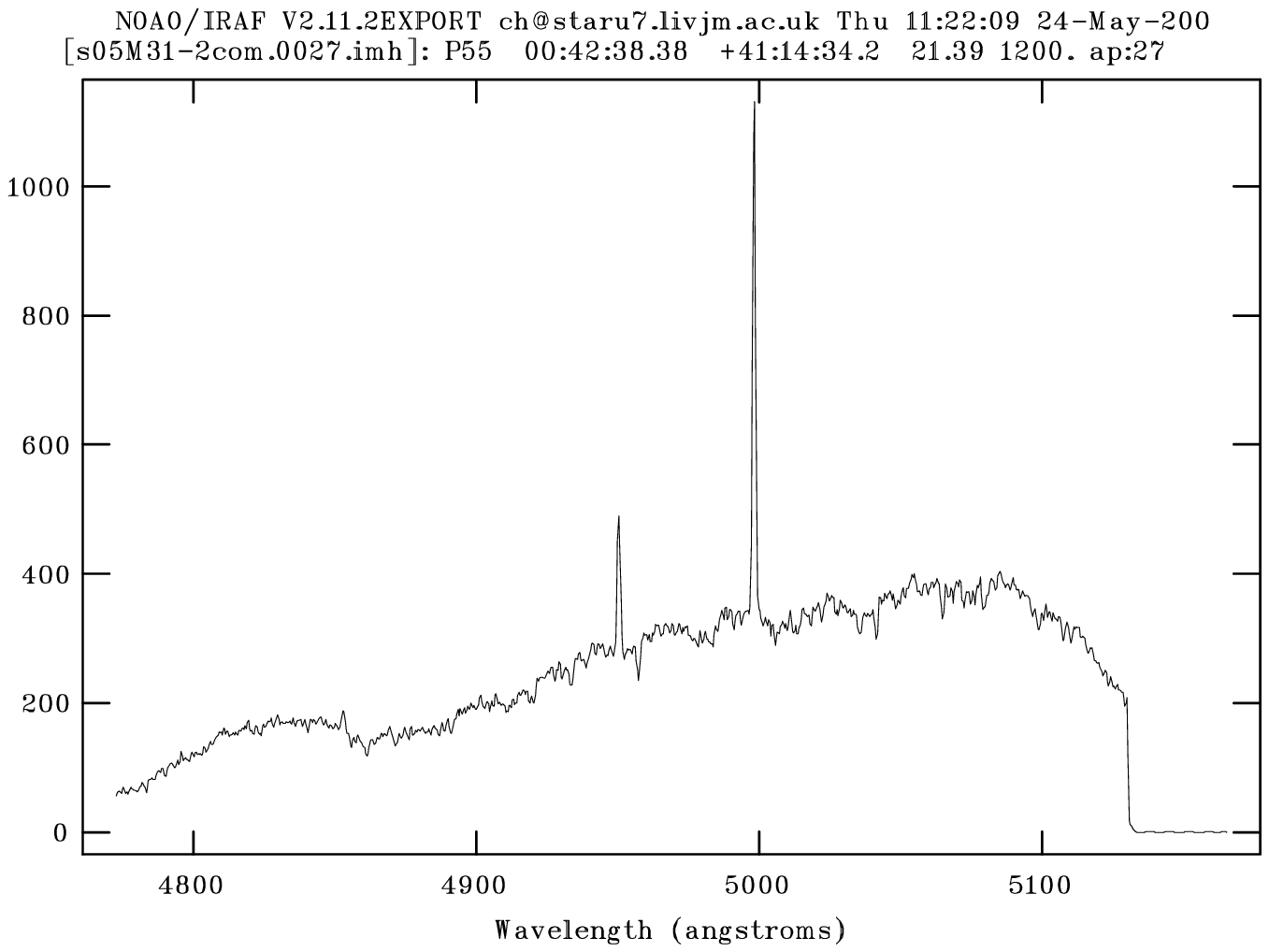,width=5.0cm,height=3.5cm}
\epsfig{file=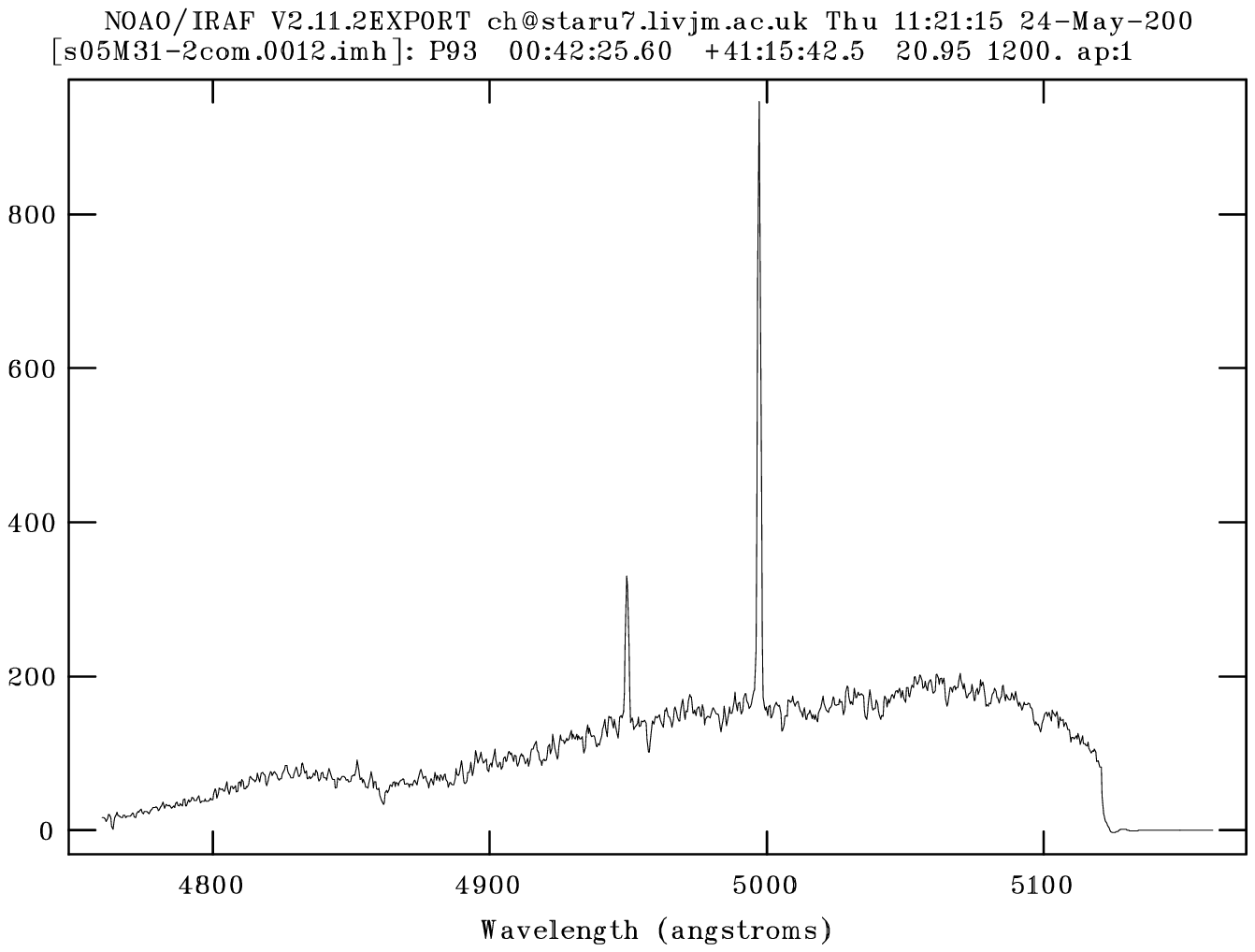,width=5.0cm,height=3.5cm}
\epsfig{file=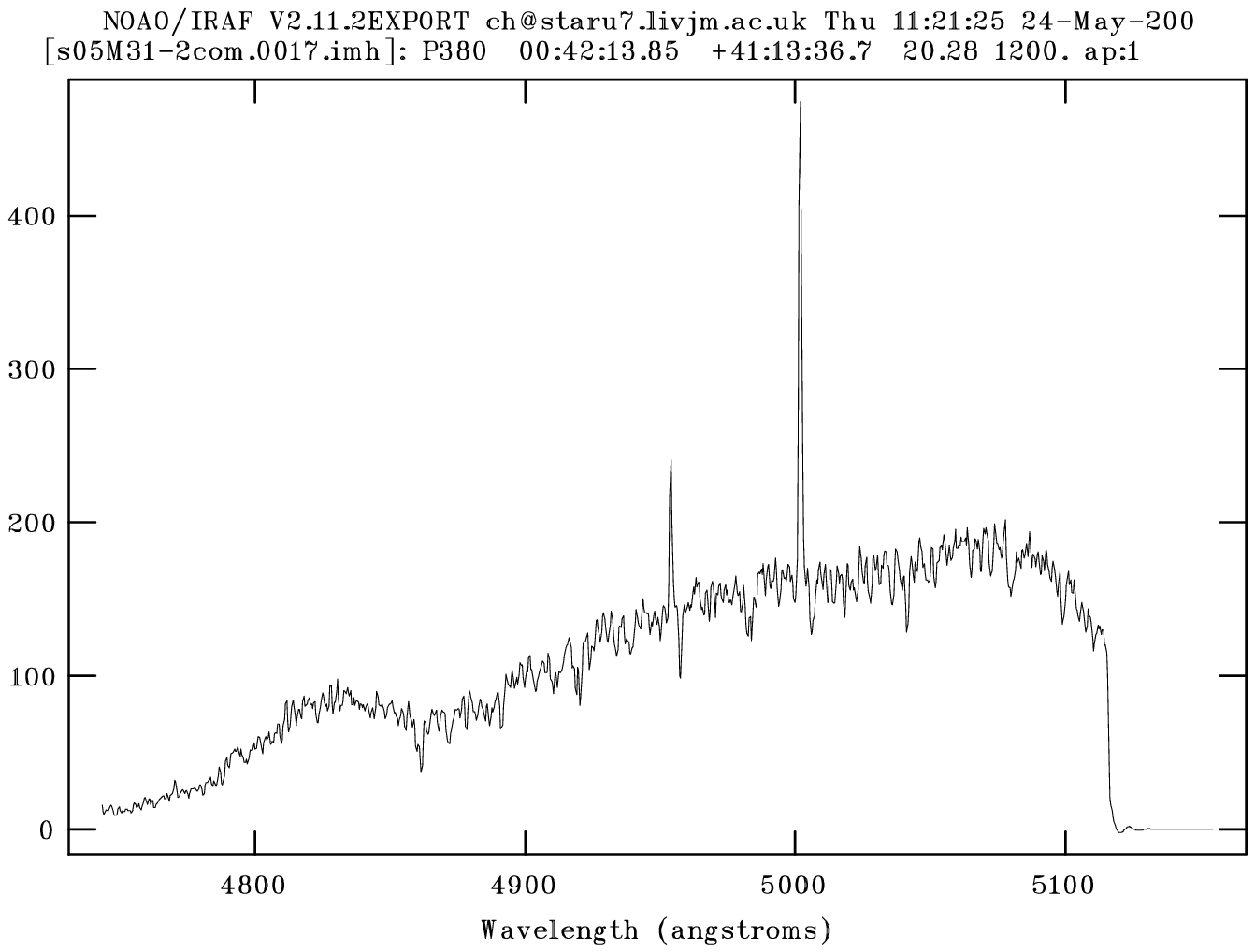,width=5.0cm,height=3.5cm}
\epsfig{file=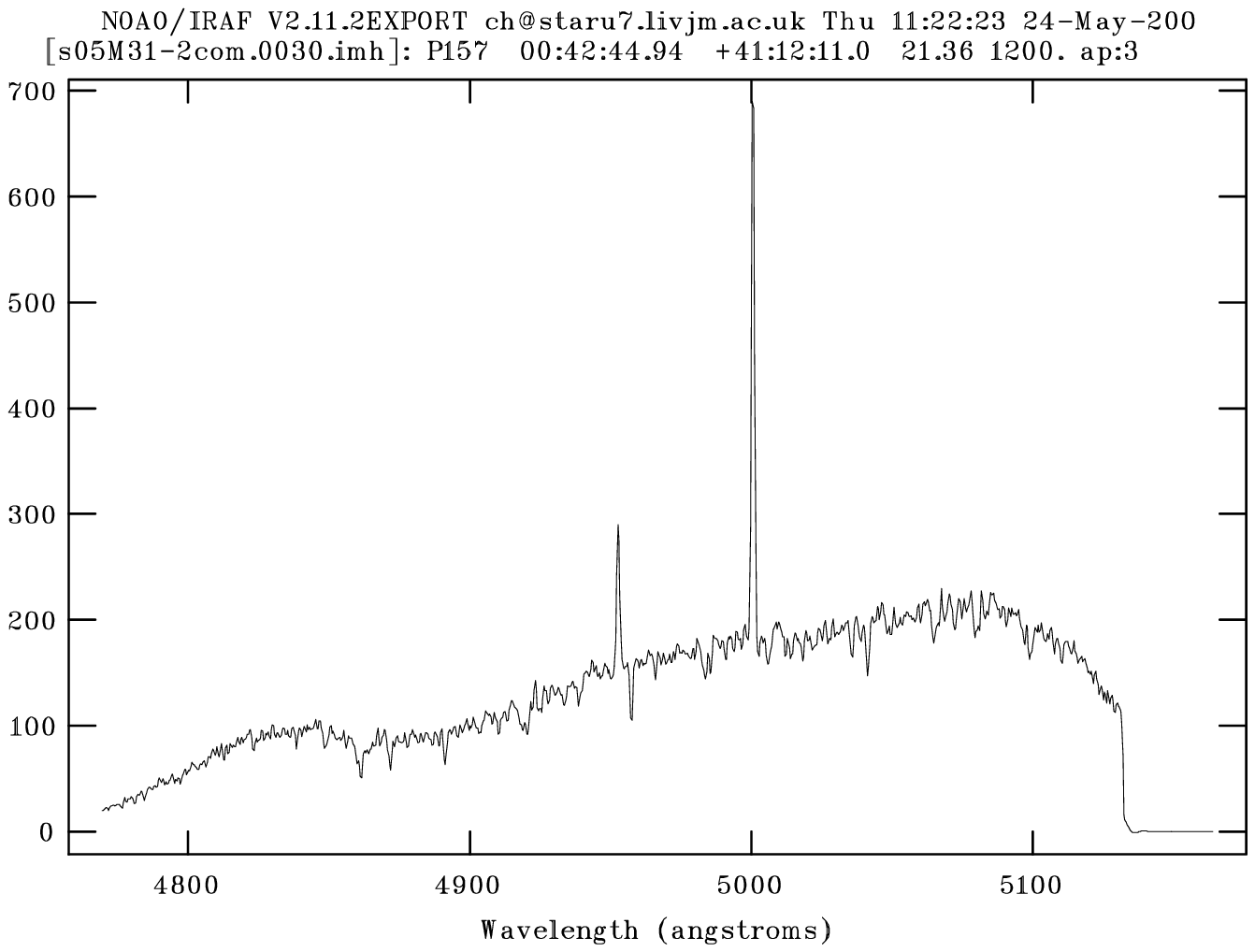,width=5.0cm,height=3.5cm}
\epsfig{file=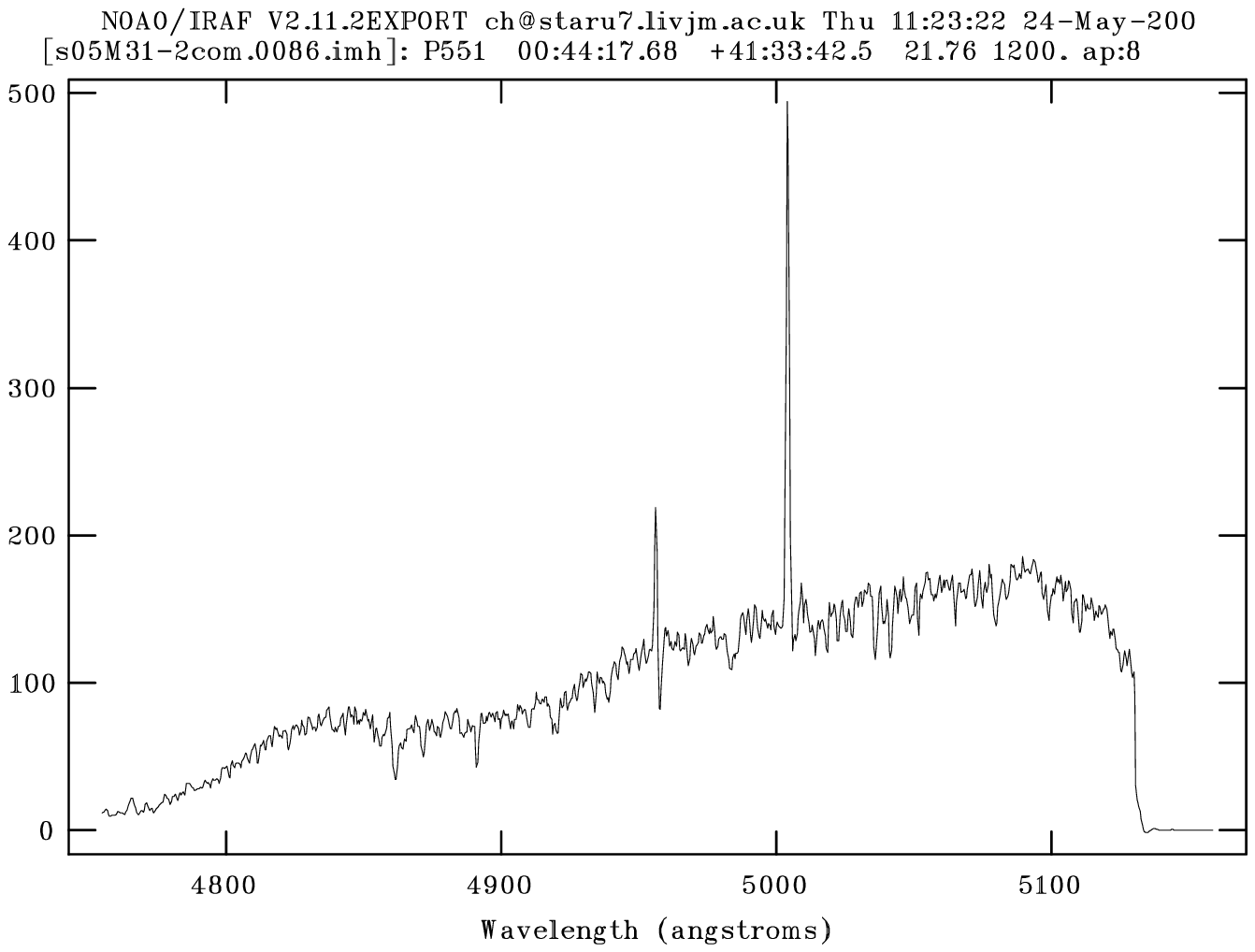,width=5.0cm,height=3.5cm}
\epsfig{file=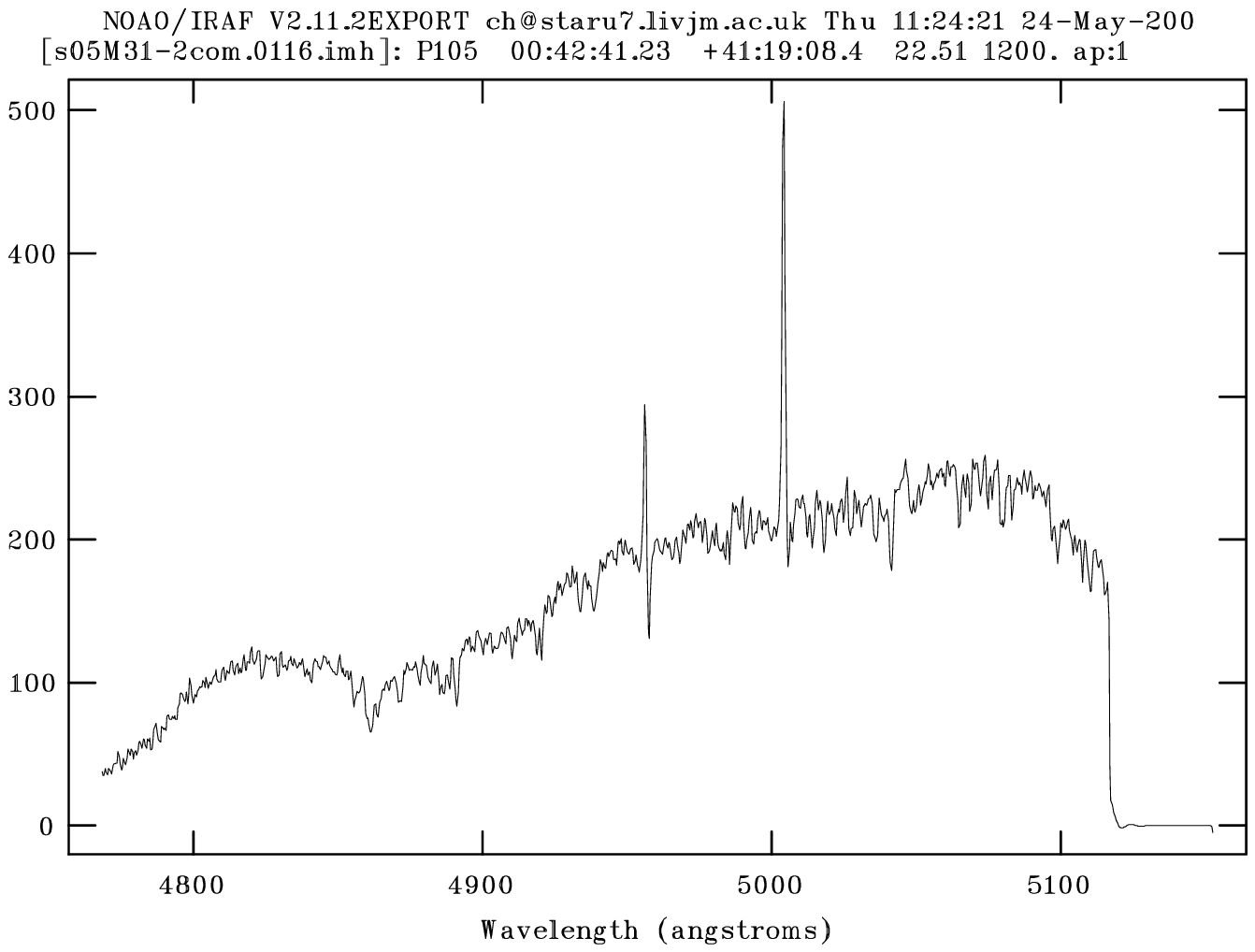,width=5.0cm,height=3.5cm}
\epsfig{file=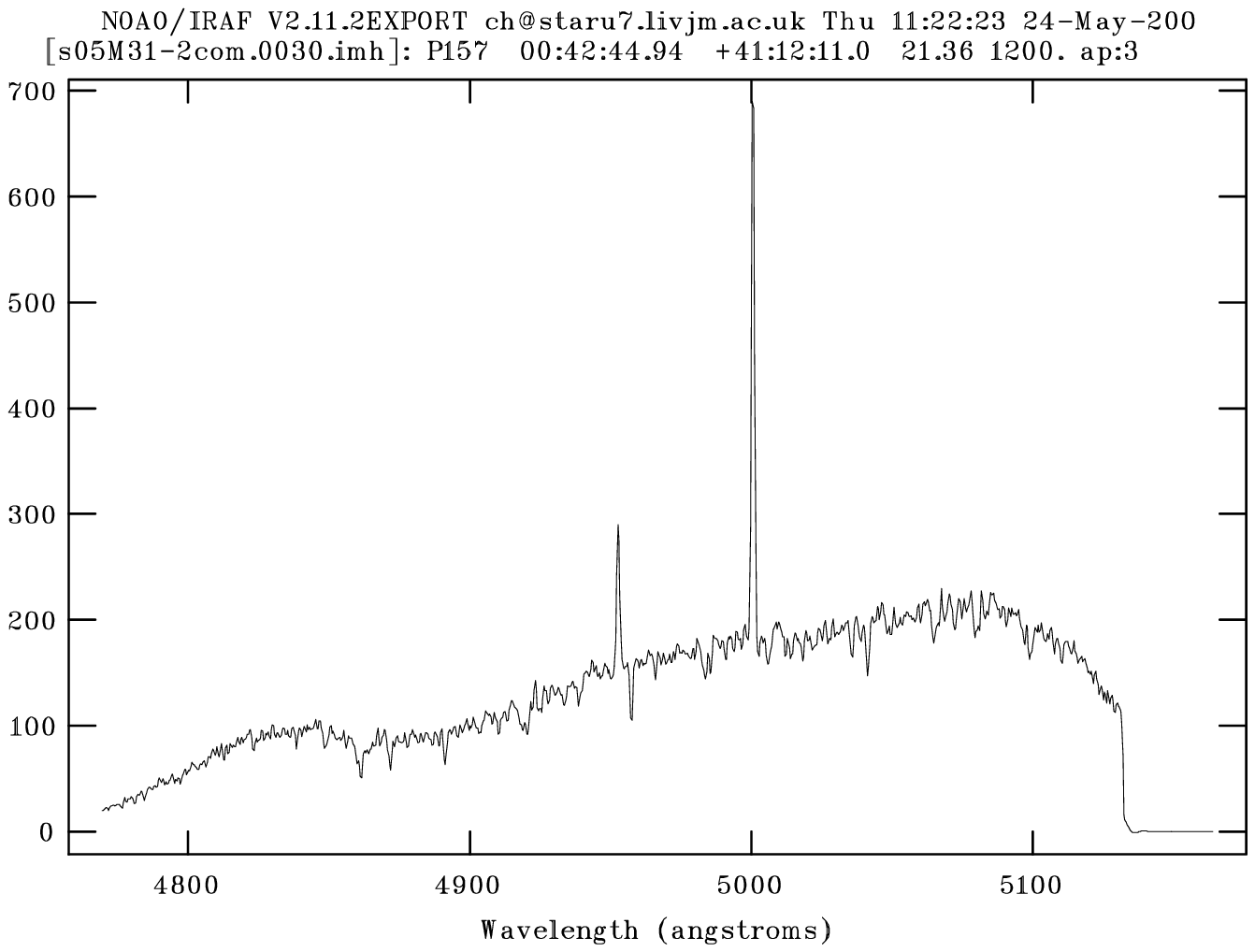,width=5.0cm,height=3.5cm}
\epsfig{file=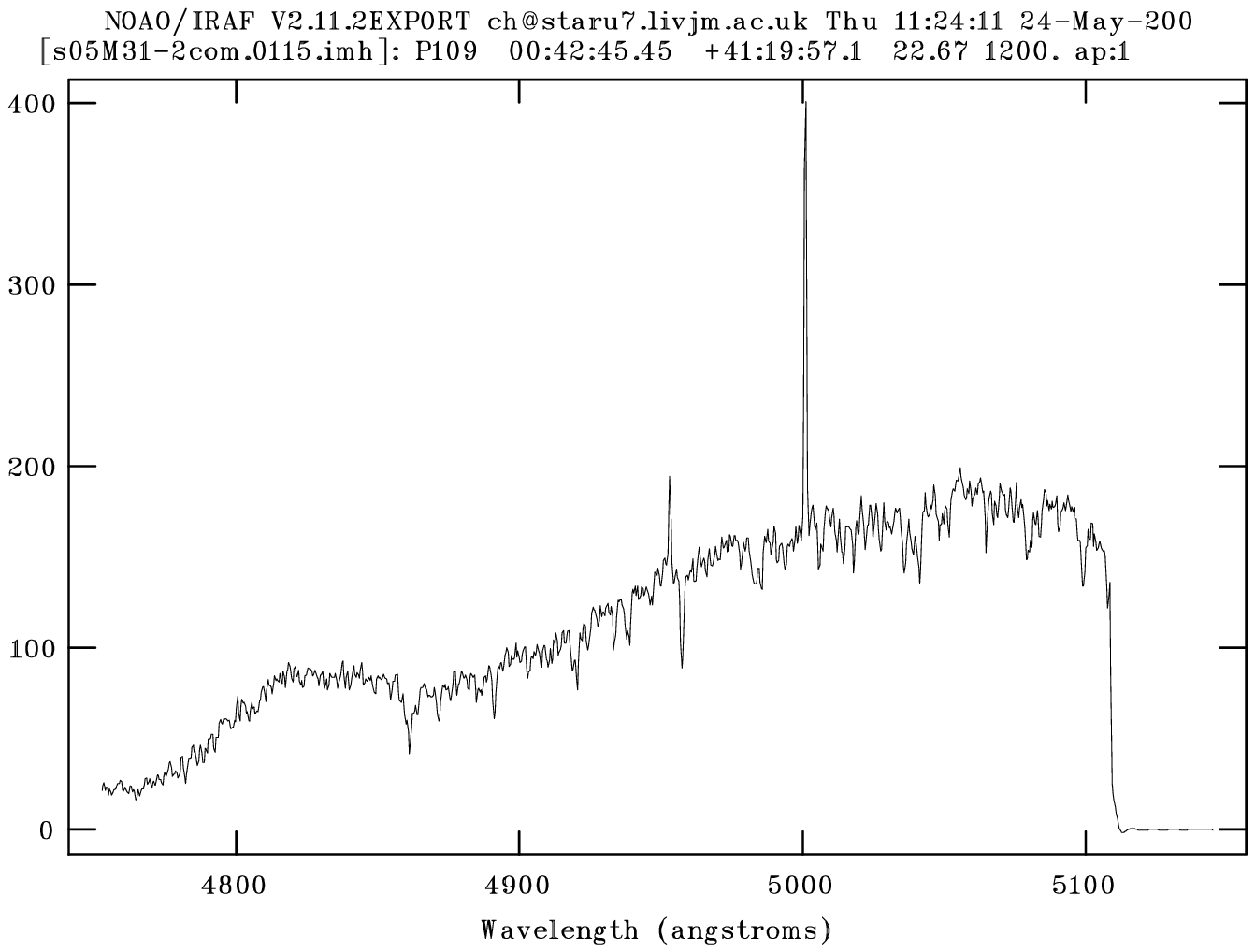,width=5.0cm,height=3.5cm}
\epsfig{file=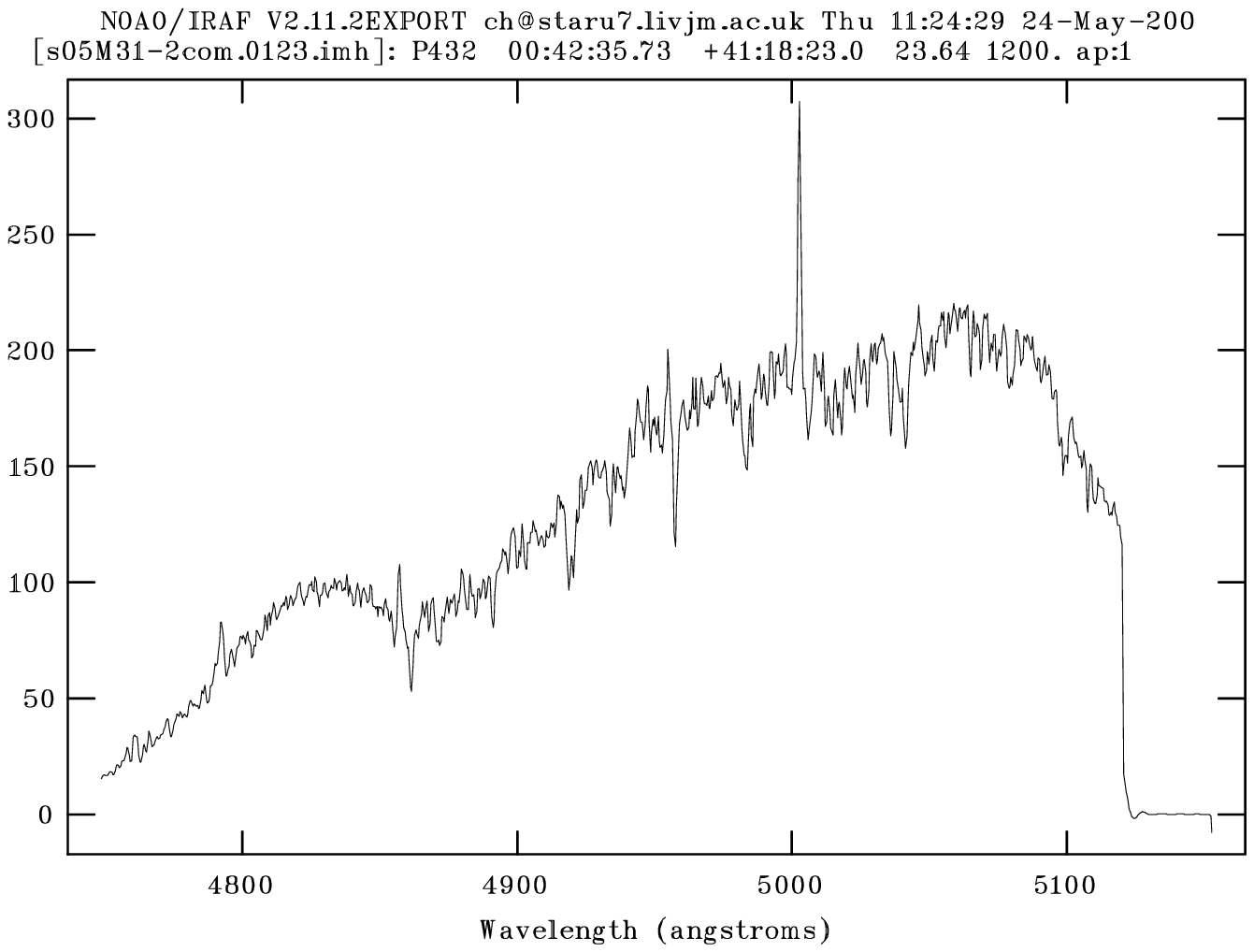,width=5.0cm,height=3.5cm}
\epsfig{file=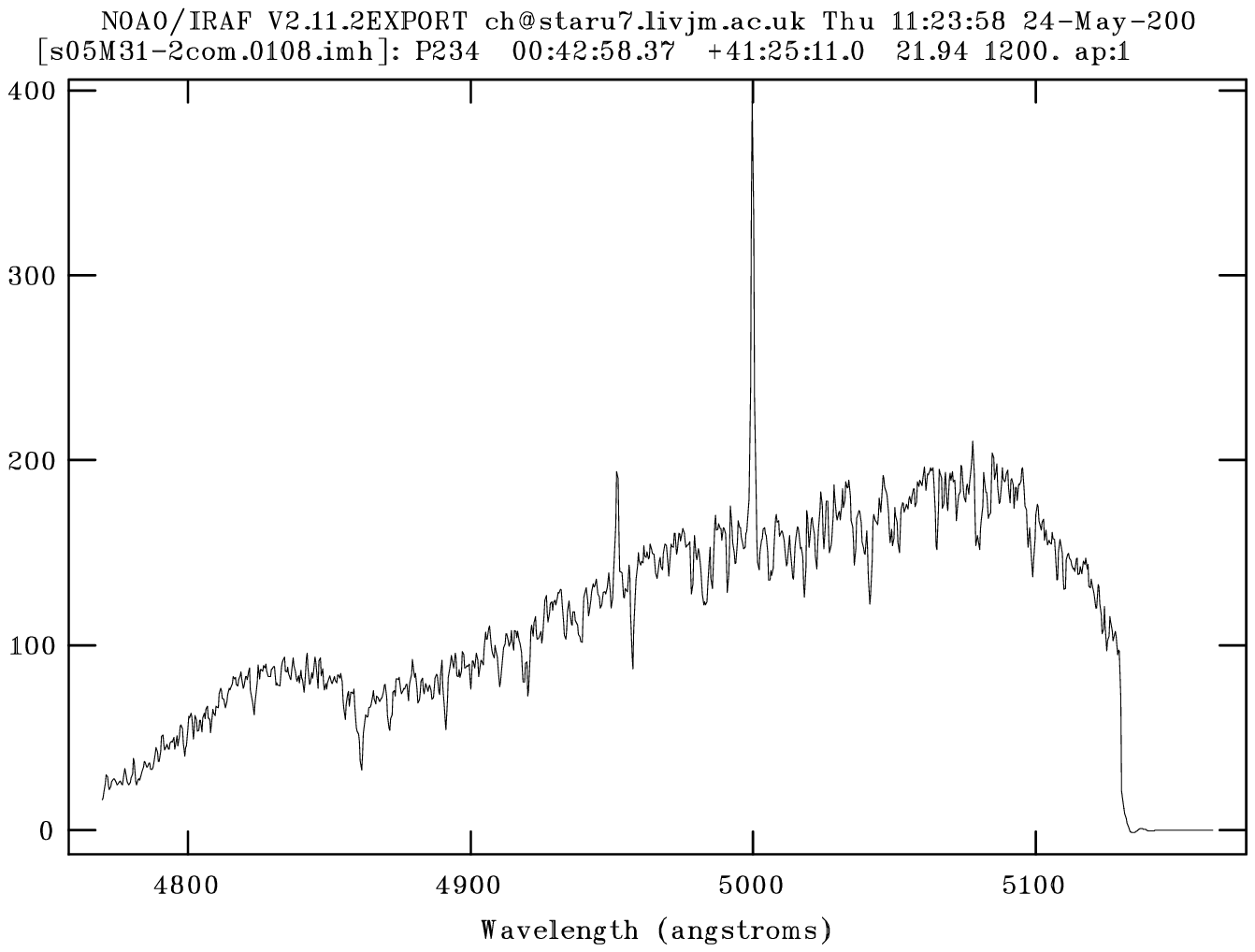,width=5.0cm,height=3.5cm}
\end{minipage} 
\caption{\label{typspec} Typical PNe spectra for combined 
exposures of different WYFFOS configuration. Both [OIII] emission
lines at 4959\AA~and 5007\AA~are visible for these spectra. The
[OIII] 5007\AA~ emission line
wavelength is used to determine the planetary nebula velocity.}
\end{center}
\end{figure*}

Data reduction was performed using the {\bf wyfred} task of the {\bf
rgo} and {\bf wyffos} packages in IRAF\footnote{Image Reduction and
Analysis Facility of the National Optical Astronomy Observatories,
Tuscon, Arizona, U.S.A..}. Aperture identification was completed using
our flat-field CCD exposures. Wavelength calibration was performed to
a typical accuracy of 0.003\AA~(0.2 km s$^{-1}$ at 5000 km s$^{-1}$):
this was achieved by fitting a polynomial of order 5-7 for
the positions of 9-12 neon arc lamp spectral lines. To optimise the
signal-to-noise ratio all exposures for a given configuration were
median-combined to produce a single frame; this also allowed a larger
number of spectra for fainter PNe to be studied. Examples of our PNe
spectra are shown in Figure ~\ref{typspec}.

Spectra were not sky subtracted. Time was not scheduled to complete offset
sky exposures during observations because sky-subtraction and flux calibration
of spectra were not essential to complete our science goals.

\subsection{Measurement of Planetary Nebulae Velocities}\label{velmeas}
\label{velocities}

A velocity was determined for each planetary nebula (PN) by 
fitting a Gaussian function to the [OIII]$\lambda$5007~and, where visible,
[OIII]$\lambda$4959~emission lines using a program provided by Dr. Bianca Poggianti 
(Dressler {\sl et al.} 1999; Poggianti {\sl et al.} 2006).

The 
central wavelength of each Gaussian was converted to a velocity
using rest-frame wavelengths in air for the lines taken from the 
NIST\footnote{National
Institute for Standards and Technology (http://physics.nist.gov/PhysRefData/ASD/index.html)} 
Atomic Spectra Database. Many
of the detected emission lines had slightly square-topped profiles
which could not be well fit by a Gaussian. To improve fitting each PN
spectrum was convolved with a Gaussian of width 0.5 pixels; this
smoothing had a negligible effect on the measurement of wavelength. 

The [OIII]$\lambda$4959 line was used only in the estimation of the errors.
This line is weaker than [OIII]$\lambda$5007 by a factor 3, and in addition
is more affected by absorption lines in the sky background spectrum;
as our spectra were taken in a mixture of grey and bright time the 
background is largely scattered solar light. 

Velocities calculated from the wavelengths of the [OIII] lines were 
corrected to heliocentric using corrections calculated using the  
Starlink program {\tt rv}. Internal velocity errors were estimated
by comparing the velocities measured with [OIII]$\lambda$4959 and [OIII]$\lambda$5007
lines. Each velocity measurement will have an error due to the error
in the estimate of the centroid of the line, and a calibration error
due to thermal or mechanical shifts between PNe and comparison lamp 
spectra, and systematic errors in wavelength fitting. The first will 
depend upon the strength of the line, and will
affect the [OIII]$\lambda$4959 and [OIII]$\lambda$5007 velocity measurements
independently, whereas we assume the second will not depend upon the
strength of the lines, and will affect the [OIII]$\lambda$4959 and 
[OIII]$\lambda$5007 velocity measurements identically in an individual
spectrum. We estimate the first from differences between the velocity
measurements of the two lines in individual spectra, and the second
by comparing velocities from repeat observations of PNe.  
%For each [OIII] emission line, a measure of equivalent width
%(hereinafter EW) was made. This is not a robust measurement because
%it measures the strength of the line not with respect to the PNe continuum,
%but with respect to the sky (the spectra are not sky subtracted) and the 
%M31 background. The EW measurement corresponds however to the detectability of the
%emission line. We assess our internal errors therefore as a function
%of EW.

We parameterise the error measurements in terms of the detectability
of the emission line, which we have chosen to quantify in terms of the ratio
of the emission line strength to the background (sky + galaxy) continuum, as the
spectra are not sky subtracted. For each [OIII] emission line, a measure of 
this ratio was made, which we express as an equivalent width (EW).  

\begin{figure}
\begin{center}
\rotatebox{270}{\includegraphics[width=6cm]{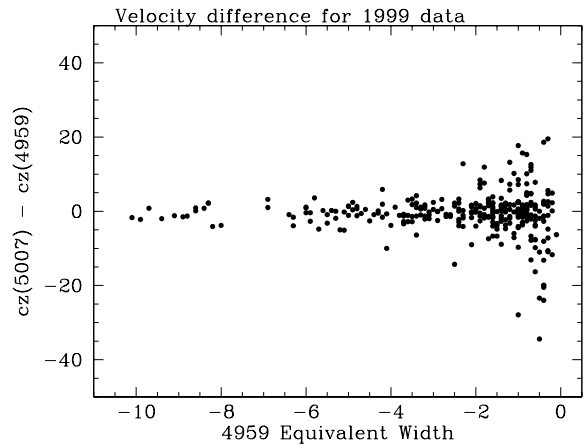}}
\end{center}
\caption{\label{comp1999}
Difference between velocities calculated from $\lambda$5007 and
$\lambda$4959, plotted against the strength of the weaker line,
for the 1999 data.
}  
\end{figure}

\begin{figure}
\begin{center}
\rotatebox{270}{\includegraphics[width=6cm]{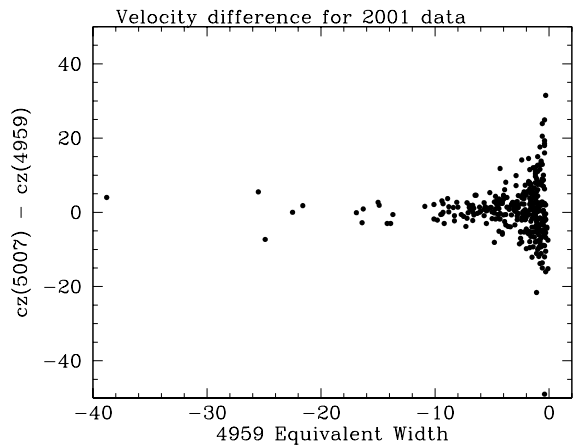}}
\end{center}
\caption{\label{comp2001}
Difference between velocities calculated from $\lambda$5007 and
$\lambda$4959, plotted against the strength of the weaker line,
for the 2001 data.
}  
\end{figure}

PNe were separated into five bins of EW measurement of the [OIII]$\lambda$4959
line. For PNe in the bin of the highest [OIII]$\lambda$4959 EW measurement, we calculated
the rms difference between the velocities calculated from 
[OIII]$\lambda$4959 and [OIII]$\lambda$5007, and assumed that  errors in the 
wavelengths of the two lines contributed equally to these differences.
This gives a value of the centroiding error for the lines in the 
strongest bin. For weaker bins, we assume that the centroiding error
on the [OIII]$\lambda$5007 line is the same as in the strongest bin, and
obtain a somewhat conservative estimate of the centroiding error for 
[OIII]$\lambda$4959 lines in each strength bin. In figures ~\ref{comp1999}
and ~\ref{comp2001} we plot the difference between the velocities 
measured from the two [OIII] lines against the ``equivalent width'', 
a measure of the strength of the [OIII]$\lambda$4959 line, for the 1999 
and 2001 runs respectively. 

Next we assume that the centroiding error on a [OIII]$\lambda$5007 line is 
the same as that on a [OIII]$\lambda$4959 line of the same ``equivalent
width''. Thus we construct a table of errors on the velocities
computed from the [OIII]$\lambda$5007 lines. 

In each run there are a number of repeat measurements, for each pair
of measurements we compute the difference. We then restrict the 
comparison to lines with [OIII]$\lambda$5007 in the top two bins by line 
strength (for which the centroiding errors are nearly equal). We 
assume that the rms velocity difference is composed of centroiding
error on each spectrum, plus calibration error, added in quadrature. 
Thus we compute the calibration error. 

The errors presented in Table ~\ref{datatable} are the sum in quadrature 
of these two components. This analysis was carried out separately on
1999 and 2001 data. Both the centroiding and calibration errors were
somewhat higher in 2001 -- in the case of the centroiding error we believe 
this to be due to brighter sky conditions.

All duplicate observations were then combined, with weights calculated
from the inverse variances, and a final error calculated from summing 
the inverse variances. Analysis of objects observed in both 1999
and 2001 showed that there was no systematic offset between the velocity
systems of the two runs, and that the rms difference was consistent with
the individual errors upon the measurements. 
%In figure ~\ref{fig:1999-2001} we plot velocities from the two runs against each other. 
At this point we
noted one very discrepant object, which we discovered to be very close to 
edge of the frame in the 2001 data, and found that the calibration procedure
had failed at the edge of this frame. This point, and points around it in
the data frame, were then rejected from the dataset, and do not appear in
the final table.

%\begin{figure}
%\begin{center}
%\rotatebox{270}{\includegraphics[width=6cm]{figure3.ps}}
%\end{center}
%\caption{\label{fig:1999-2001}
%Velocities of objects observed in both runs: 1999 velocity (X axis)
%plotted against 2001 velocity (Y axis).
%}  
%\end{figure}

Our final velocities are listed 
in Table ~\ref{datatable}. In this table the first column gives an identifier; 
if it is listed by C89 their identifier is used (of the form Pxxx), otherwise
it has an identifier from our imaging survey (PN\_field\_object). Columns
2 and 3 give the position in J2000 co-ordinates, column 4 and 5 the final 
heliocentric velocity and its error. 
Columns 6 and 7 give the distance from the nucleus along the major axis and 
minor axis respectively, where we have adopted the position of the nucleus
listed in the NASA NED database, close to that given by Crane {\sl et al.} (1992).
We assume a major axis position angle of 38.1$^{\circ}$. Column 8 gives the distance from the
nucleus, again in arcminutes, and column 9 the azimuthal angle (measured through
west from north) in degrees. Column 10 gives the number of observations 
combined to give the final velocity.

HK04 published a list of 135 PNe velocities
in fields to the South and East of the centre of M31. Some of their area 
lies outside our survey area, but there are 42 targets in common. 
Comparison of these results with ours shows that there are two objects
whose velocities are discrepant. For 
PN\_5\_1\_3 (=HKPN121) we have detections of [OIII]$\lambda$5007 and [OIII]$\lambda$4959
at the same velocity; this is in the highest error bin of HK04, and we 
are confident that our velocity is correct. For
PN\_2\_1\_15 (=HKPN66) there is no clear reason for the discrepancy, but
we do not see [OIII]$\lambda$4959 or H$\beta$, so this could be a misidentification
of the line on our part. 

Neglecting these discrepancies we find a mean velocity difference of 8.7 km/s,
in the sense that the HK04  velocity system is positive
with respect to ours, and an rms velocity difference of 10.2 km/s after
subtracting the zero point difference, which is 
consistent with the quoted errors on the two datasets. 
%In Figure ~\ref{HKcomp} we plot the HK04 velocities against ours.

%\begin{figure}
%\begin{center}
%\rotatebox{270}{\includegraphics[width=6cm]{figure4.ps}}
%\end{center}
%\caption{\label{HKcomp}
%Velocities from HK04 (Y axis), plotted against those from this paper
%(X axis).
%}  
%\end{figure}

Merrett {\sl et al.} (2006) measure velocities for almost all of our PNe, 
and find a systematic offset and trend, which they attribute to a systematic shift
in the PNS wavelength calibration. After transforming their velocities to the
system of the current paper, they find a combined error 
$\sigma = (\sigma_{PNS}^2 + \sigma_{H06}^2)^{1/2}$ = 15.4 km/s. This is 
consistent with the errors quoted in the present paper, and the expectation for the 
mean errors in the PNS velocities.

\begin{figure}
\begin{center}
\rotatebox{270}{\includegraphics[width=14cm,height=14cm]{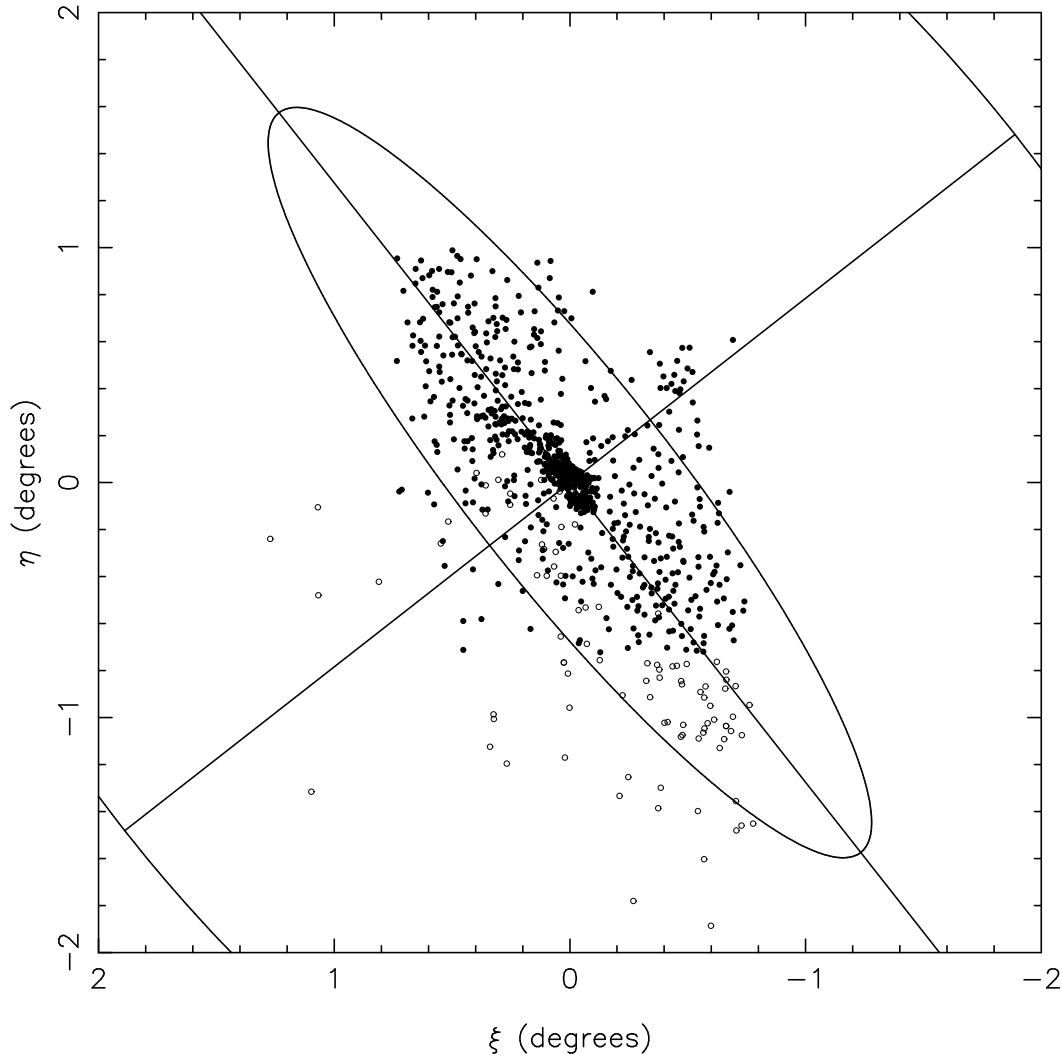}}
\end{center}
\caption{\label{Positionplot}
Positions of PNe with measured velocities, relative to the
centre and orientation of M31. Filled symbols represent PNe from the
present paper, open symbols PNe for HK04 which are not common with our
sample. The inner and outer ellipses are drawn using a position angle
of 38.1$^{\circ}$. The outer ellipse represents a flattened ellipsoidal
halo (aspect ratio 3:5) of semi-major axis radius 4$^{\circ}$ ($\approx$
55~kpc).  The inner ellipse has a semi-major axis of 2$^{\circ}$
($\approx 27$~kpc) and represents an inclined disk with
$i=77.5^{\circ}$. 
}  
\end{figure}

In Figure ~\ref{Positionplot} we show the positions of the PNe in our sample using a
standard coordinate projection centred on M31.  The ellipses overlaid in
the plot represent the disk and a flattened halo.

\section{Kinematics of the PN Data Set}
\label{analysis}

In this section we present a brief analysis of the kinematics of our
PN data set. Full dynamical modelling of these data, along with a
number of other new kinematic data sets in M31 will be presented
elsewhere. In the following, we use the convention that the positive
$X$-axis is along the receding side of the major axis of M31. The
transformations between RA and Dec and ($X,Y$)
are as given by Huchra {\sl et al.} (1991). We assume that the centre of M31
is at (R.A., Dec.) = 00:42:44.324, +41:16:08.53 (J2000) and that the
position angle of the major axis is $38.1^{\circ}$ (Ferguson {\sl et al.} 2002).

\begin{figure}
\begin{center}
\includegraphics{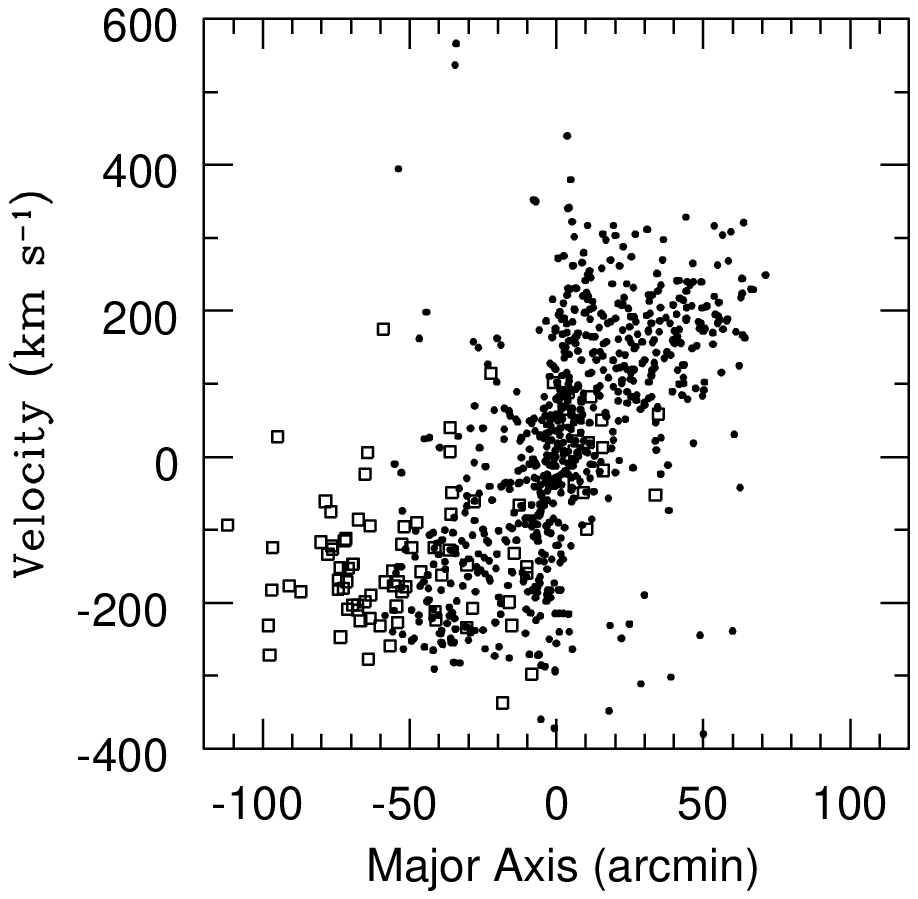}
\end{center}
\caption{
PN velocity versus distance along the major axis for our
complete data set (solid symbols). The open squares denote those PNe
observed by Hurley-Keller et al. (2004) which were not re-observed as
part of our survey.
}
\label{fig:V_vs_X_all}
\end{figure}

Figure~\ref{fig:V_vs_X_all} plots the velocities of our PNe versus
their major axis position. From the Figure, it is clear that our data
set is dominated by rotationally supported components, as would be
expected given the spatial distribution of our targets, which mostly
probe the disk and bulge of M31.

\begin{figure}
\begin{center}
\includegraphics{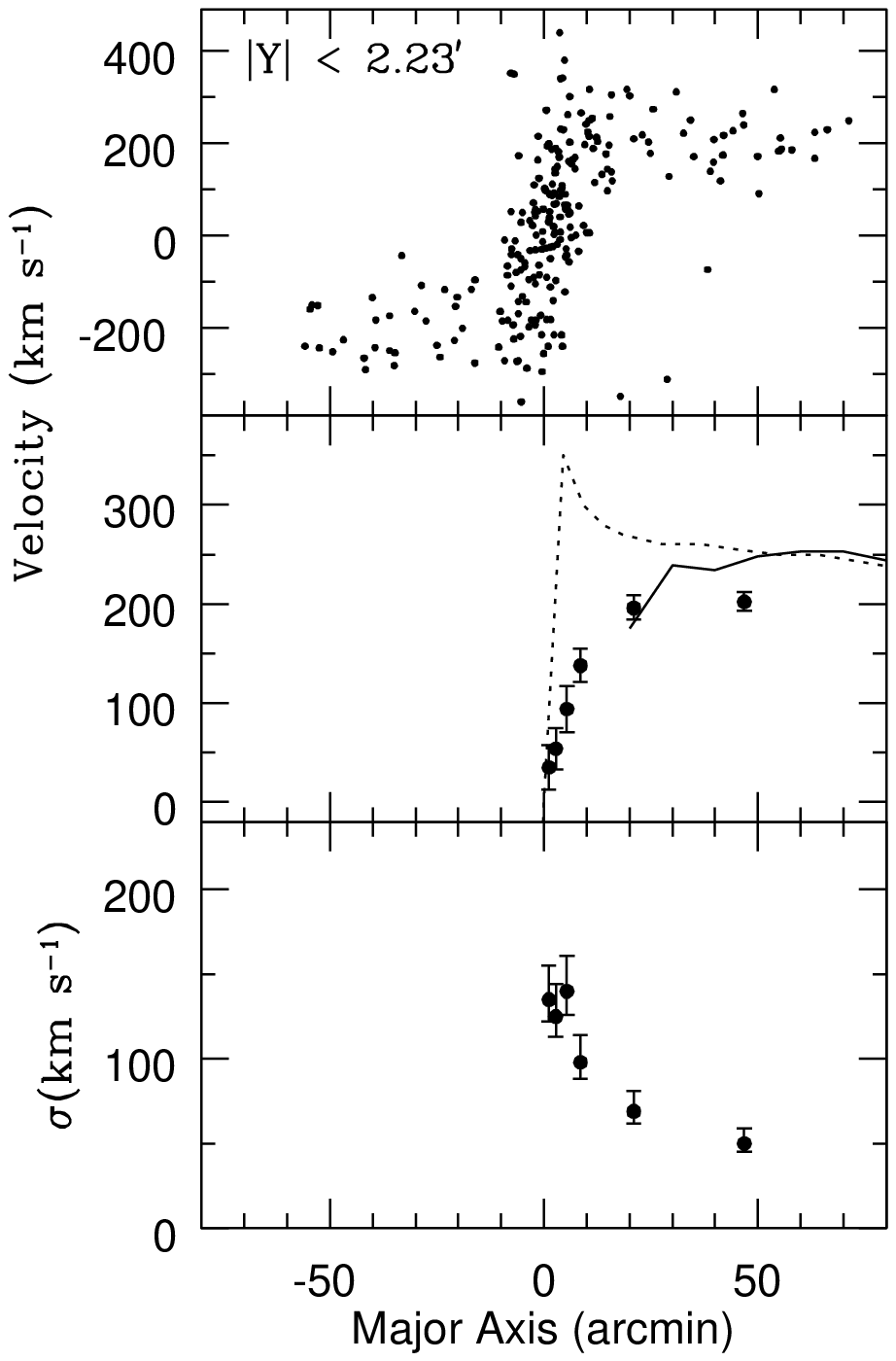}
\end{center}
\caption{Line-of-sight velocity distribution for PNe with $|Y|< 2.23^\prime$ (0.5 kpc). The top
panel shows the PN velocity versus distance along the major axis. The
middle and lower panels show, respectively, the mean velocity and
velocity dispersion (with $1\sigma$ errors) as a function of major
axis position. The sold curve in the middle panel is the gas rotation
curve of Kent (1989) converted to line-of-sight velocity for comparison with
our data; the broken curve is the rotation
curve of Braun (1991). The data have been folded about the
minor axis. See text for a detailed discussion.}
\label{fig:V_vs_X_inner}
\end{figure}

Simple geometric arguments imply that for a star with minor axis
distance $|Y| < 0.5$kpc, the line of sight velocity is close to the
azimuthal velocity. By considering stars close to the major axis we can
therefore obtain a meaningful stellar rotation curve and line of sight
velocity dispersion profile. The top panel of
Figure~\ref{fig:V_vs_X_inner} displays the velocities against major
axis position for PNe with $|Y|<0.5$kpc (or $2.23^\prime$). The middle and
bottom panels of Figure~\ref{fig:V_vs_X_inner} show the velocity and
dispersion profiles that we determine from the data.  In order to fit
the velocity and dispersion profile we assume that at each point the
underlying velocity distribution follows a Gaussian with mean velocity
$\bar{v}(r)$ and dispersion $\sigma(r)$. We fold the data assuming a
systemic velocity for M31 of $-310$km\,s$^{-1}$ and construct radial
bins containing 40 PNe
\footnote{The outer bin contains $21$ objects.}each with velocities
$v_{i}$ and corresponding measurement errors $\sigma_{i}$. The probability of obtaining the
ensemble of observed velocities in a radial bin is:

\begin{equation}
{P(\{ v_{i}\}|\{\sigma_{i}\},\bar{v},\sigma)}=\prod_{i}\frac{1}{\sqrt{2\pi( \sigma_{i}^{2}+\sigma^2)}}
\exp\left(-\frac{(v_{i}-\bar{v})^{2}}{2(\sigma_{i}^2+\sigma^{2})}\right)
\end{equation}   

Equation (1) is only strictly true within a bin if the rotation
velocity is constant in the bin. In the inner zone where rotation velocity 
is varying rapidly as a function of radius
the derived dispersion is artificially inflated by the variation of the
rotation velocity. However this is not a large effect,
as, for a uniform distribution within the
bin, the dispersion is only inflated in quadrature by the bin velocity
range divided by $\sqrt{12}$. For a velocity range
of 50 km s$^{-1}$, this results in an increase in the dispersion of about 1
km s$^{-1}$.

\begin{table}
\caption{Line of sight velocity and velocity dispersion data (plotted in
Fig.~\ref{fig:V_vs_X_inner}). Columns
give: (1) major axis position $X$ in arcmin; (2) mean velocity in
km\,s$^{-1}$; (3) miniumum and (4) maximum values of the mean velocity
($1\sigma$ range); (5) velocity dispersion in km\,s$^{-1}$; (6)
minimum and (7) maximum values of the velocity dispersion ($1\sigma$
range); (8) number of stars in the bin.}
\begin{center}
\begin{tabular}{lccccccc}
$X$ (arcmin) & $\overline{v}$ & $\overline{v}_{\rm min}$ &
$\overline{v}_{\rm max}$ & $\sigma$ & $\sigma_{\rm min}$ & $\sigma_{\rm max}$ & N\\ \hline
1.15 & 35 &  13 & 58   & 135 & 122 & 155  & 40 \\
2.79 & 54 &  33 & 75   & 125 & 113 & 144 & 39 \\
5.34 & 94 &  71 & 117   & 140 & 126 & 161 & 40 \\
8.52 & 138&  121 & 155   &  98 & 88 & 114 & 38 \\
21.00 & 196 &  184 & 209 &   69 & 62 & 81 & 38 \\
46.81 & 202 &  193 & 212 &   50 & 45 & 59 & 33 \\ \hline
\end{tabular}
\end{center}
\label{tab:V_vs_X}
\end{table}

For each radial bin we estimate $\bar{v}$ and $\sigma$ by maximising
the expression for the probability with respect to these parameters.
We account for the effect of outliers by repeating the maximisation
procedure twice.  In the first iteration we omit PN data falling more
than $2\sigma$ away from the mean velocity of the bin as defined by
the initial fitted parameters.  The second iteration uses a $3\sigma$
cut based on the updated parameters. The middle and bottom panels of
Figure~\ref{fig:V_vs_X_inner} show the results of this procedure, 
which are also tabulated in Table \ref{tab:V_vs_X}. The
errorbars in each parameter shown in the plots are $1\sigma$ errors
obtained by numerically determining the probability density function
for that parameter by marginalising over the other parameter. We then
define the $1\sigma$ confidence interval as the interval centred
around the peak of the probability density function which contains
$68$ per cent of the probability. The systemic velocity used in
folding the data is estimated using the complete data set from
Figure~\ref{fig:V_vs_X_all}. Using these data a rotation curve is
calculated, as discussed above - the systemic velocity is then
calculated by linear interpolation between the two bins on either side
of the minor axis. In this
way we obtain a value of $-310 km s^{-1}$ for the systemic velocity of M31.

It is clear from the top two panels of Figure~\ref{fig:V_vs_X_inner}
that the disk of M31 dominates the velocity distribution at these
small distances from the major axis. The profiles in the two lower
panels therefore constitute the first complete rotation curve and
dispersion profile to 50 arcmin radius for the disk of M31 based on stellar
kinematics. There have been many stellar kinematic studies in the very central regions, 
close to the black hole (e.g. van der Marel {\sl et al.} 1994). At intermediate radii
the most comprehensive study is by McElroy
(1983), who presented long-slit velocity and velocity dispersion curves
at several position angles, out to a maximum of 10 arcminutes radius.
At larger radii, the data of HK04 were
insufficient to determine the profiles of kinematic parameters. For
comparison, in the middle panel we show the gas rotation curves of
Braun (1991) and Kent (1989). Our PNe rotation curve is compatible with 
the curve of Kent (1989), however inside 30 arcminutes Braun's curve shows 
more rapid rotation than the PNe, and indeed than Kent's gas rotation curve. 
The differences between the two HI rotation curves are due largely to 
differences between the models used to turn observed gas velocity into 
circular speed. McElroy (1983) has also found the stellar velocity 
to be lower than the gas velocity, even after allowing for projection
effects. The stellar rotation is expected to be lower than
due to asymmetric drift (Gerssen {\sl et al.} 2000), and possibly also to 
inclusion of thick disk and bulge components.

\begin{figure}
\begin{center}
\includegraphics{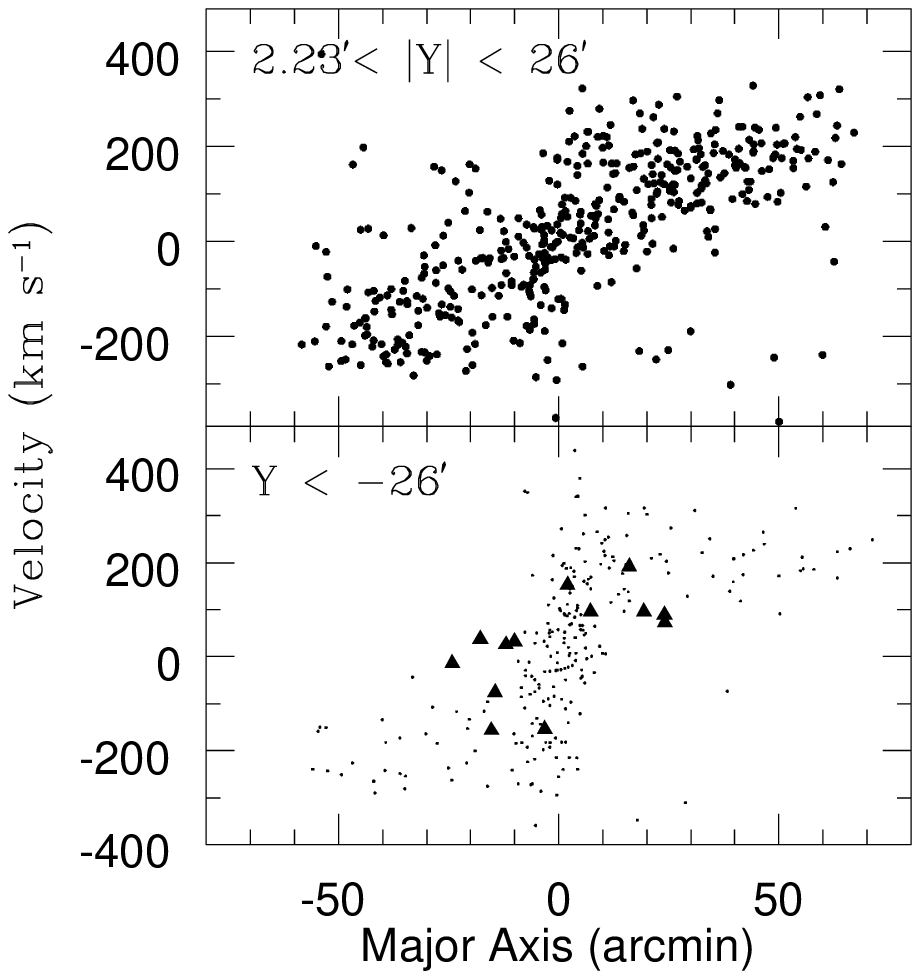}
\end{center}
\caption{
Line-of-sight velocity distributions for PNe with $2.23^\prime 
(0.5 kpc) < |Y| < 26^\prime$ (5.8 kpc) (upper
panel) and $Y < -26^\prime$ (lower panel). In the lower panel, the data
with $Y < 26^\prime$ are shown as triangles, and the data with $|Y| < 2.23^\prime$
are also plotted for comparison. See text for a discussion.}
\label{fig:V_vs_X_med}
\end{figure}

Following HK04, we illustrate the spatial variation of
the velocity distribution in our sample by dividing our data according
to minor axis distance (Figure~\ref{fig:V_vs_X_med}). Projection
effects make the velocity distributions at larger distances from the
major axis more difficult to interpret. For $|Y|>2.23^\prime$ we expect to
see a mixture of bulge and disk PNe (and potentially halo PNe
also). Moreover there will be an apparent
flattening of the rotation curve at large minor axis displacements even
for the thin disk PNe, simply because of the projection of velocity and 
position in a tilted, thin disk with circular orbits.  
The upper panel of Figure~\ref{fig:V_vs_X_med} plots the data
for $2.23^\prime <|Y|< 26^\prime$ (the optical edge 
of the disk projects to
approximately $5.8 kpc$ or $26^\prime$along the minor axis).  It is clear that the
PNe in this sample are rotationally supported - the greater spread in
velocities for a given $X$ (compared to Figure~\ref{fig:V_vs_X_inner})
is due to the larger range of physical radii probed by these stars.

In the lower panel of Figure~\ref{fig:V_vs_X_med} we plot PNe with
$Y < -26^\prime$ as solid triangles, with the data with $|Y| < 2.23^\prime$ also
shown for comparison; we omit PNe with $Y > 26^\prime$ as our sample in
that region is dominated by likely members of the M31 satellite NGC
205 (based on their velocities and positions).  The triangles in this
plot show a similar trend to the data in the upper panel. This suggests
that our data beyond the edge of the projected disk are still
rotationally supported.  This is consistent with the claim of
HK04 that their sample of PNe outside $5.8$kpc belong
to an extended bulge rather than to a halo. Although our data at these
large radii are clearly limited, no obvious signature of a hot,
pressure-supported halo population is evident in the data set.

\begin{figure}
\begin{center}
\includegraphics{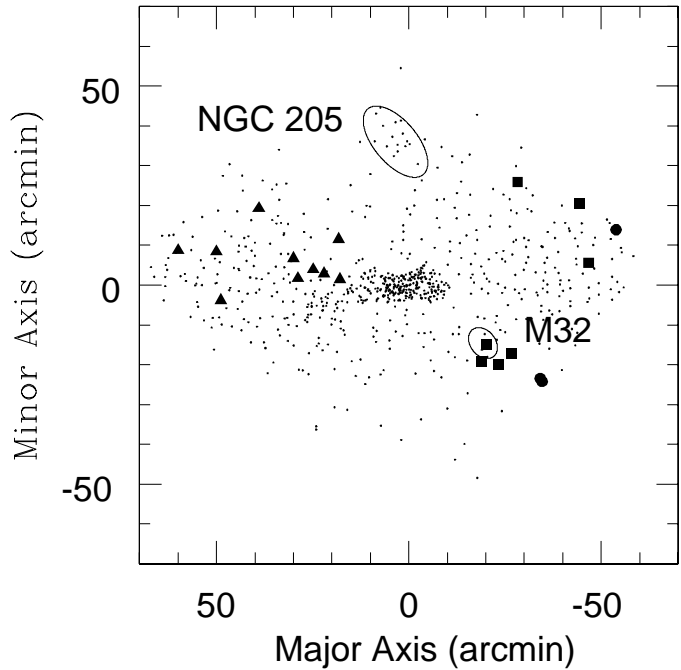}
\end{center}
\caption{
The spatial distribution of velocity outliers from
Figure~\protect\ref{fig:V_vs_X_all} compared with the distribution of
our full sample (points). PNe with $X>10^\prime$ and $V <
-150$km\,s$^{-1}$ are shown as black triangles; those with
$X<-10^\prime$ and $V > 100$km\,s$^{-1}$ are shown as black
squares while those with $X<-10^\prime$ and $V >
300$km\,s$^{-1}$ are shown as black discs. The positions of M32 and
NGC205 are indicated - their angular sizes and position angles are
taken from de Vaucouleurs et al. (1991).}
\label{fig:Outliers}
\end{figure}

One of the main goals of kinematic studies of the disk and bulge
regions of M31 is to establish the evolutionary history of the various
components (see Morrison {\sl et al.} 2004 for a detailed discussion). In
this context, kinematic substructures are of particular interest as
they may provide clues to the relative roles of mergers, accretions and
in situ star formation. A number of recent studies have found evidence
of substructure in M31, both kinematic and
spatial (e.g. Ibata {\sl et al.} 2001; Ibata {\sl et al.} 2005; Merrett 
{\sl et al.} 2003). In
Figure~\ref{fig:V_vs_X_all} there are a number of conspicuous velocity
outliers and in Figure~\ref{fig:Outliers} we show the spatial
distribution of those PNe with $(X>10^\prime,V<-150$km\,s$^{-1})$ and
$(X<-10^\prime,V>100$km\,s$^{-1})$ whose velocities imply that they
are counter-rotating relative to the main disk population. We note
that, based on their velocities and positions, the group of PNe
located near $(-20,-20)$ are very likely to be associated with the
satellite galaxy M32. Another group of PNe at 
(5,40) are likely associated with NGC 205. 
A third group of PNe in the vicinity of
$(25,0)$ appears to belong to a counter-rotating population that is
not associated with a known satellite. 
These PNe have velocities consistent with possible membership of the giant
southern stream (Ibata et al. 2001; 2004) since after traversing the
nuclear region, the general symmetry of the potential field implies that
far-side stream members will have similar velocities to those seen at the
near-end of the stream on the south side of M31.
Merrett {\sl et al.} (2003) find a similar counter-rotating
population of PNe, which appears to connect the southern stream to the
northern spur, and five of the outliers in Figure \ref{fig:Outliers}
are common with their candidate stream extension objects. 
One or two groups of globular clusters were also found at this location by  
Perrett et al. (2003), reinforcing the identification of this component.
It is also possible that the other small group of
outliers with $X<0, Y>0$ could be part of the stream detached on
a previous passage. We will investigate further possible substructure
in the context of a complete dynamical model of M31 in a later paper.
 
\section{Conclusions}

PNe velocities provide an important tool for investigating the detailed kinematics
of galaxies, and we have shown that we can measure the velocities of 
large samples of PNe to an accuracy of 6-10 km s$^{-1}$. From our velocities
in M31 we show that the stellar rotation curve agrees in the outer region with
the published HI rotation curves, with a rotation velocity of $\approx$ 200 km s$^{-1}$
at our outer limit of 50 arcmin (12 kpc).
In the inner region the PNe show somewhat less rotation than the gas,
in agreement with the predictions of asymmetric drift models. Figure 8 shows that the 
PNe velocity dispersion drops from $\sim$ 130 
km/s at the galaxy center to $\sim$ 50 km/s at 50 arcmin ($\sim$ 11 kpc) 
along the major axis. As noted earlier, and in agreement with HK04, we 
find no evidence for 
a dynamically hot PNe halo with a significant velocity dispersion. This is 
in contrast to the M31 globular clusters, which have a global velocity 
dispersion of $\sim$ 150 km/s, larger than that of the PNe (Perrett et al. 
2002; unfortunately Perrett et al. do not present the radial profile for 
the M31 globular cluster velocity dispersion). The M31 globular clusters 
have a rotation velocity of $\sim$ 140 km/s, lower than found here for the 
disk PNe ($\sim$ 200 km/s), though the metal-rich clusters have a 
higher rotation velocity of 160 $\pm$ 20 km/s.
We provide confirmation of the kinematic subpopulation, counter-rotating with
respect to the disk and the main bulge, found by Merrett {\sl et al.} (2003)
and identified with the extension through the plane of the southern stream of M31.
Future analysis of the current dataset, combined with those of Merrett
{\sl et al.} (2006) and HK04 promise to yield a much more complete kinematic, dynamical
and mass model of M31.

\section*{Acknowledgments}

We thank D. Clarke for providing references for Solar
absorption lines, C.A. Collins for helpful discussions on error
analysis, and Bianca Poggianti for providing the software for 
measuring the line wavelengths.
We are grateful to Heather Morrison for providing the data for the
Braun (1991) and Kent (1989) rotation curves in electronic form.
MIW acknowledges PPARC for funding. DPQ thanks the National University
of Ireland and EPSRC for financial support. We thank an anonymous referee
for very helpful comments on the original draft of the paper.

The William Herschel Telescope and the Isaac Newton Telescope
are operated on the
island of La Palma by the Isaac Newton Group in the Spanish
Observatorio del Roque de los Muchachos of the Instituto de
Astrof\'{i}sica de Canarias. The authors acknowledge the use of 
Starlink facilities in the reduction of the data.

\clearpage

\appendix

\section[]{Planetary Nebulae Velocity Measurements}

\begin{longtable}[c]{lllrrrrrrr}
\caption{\label{datatable}
Positions and Velocities of Planetary Nebulae.}

\endhead

\endfoot

\hline 
Identifier	&	RA   &	Dec       &   cz     &	error & Major axis &Minor Axis&Radius   &Azimuth &N	\\	
                &    (J2000) &    (J2000) &   (km/s) &  (km/s)& Arcmin     & Arcmin   & Arcmin  & Degrees&Obs  \\ \hline	
PN\_8\_1\_1 & 00 38 43.65 & 41 14 11.2 & -271.0 & 8.2 & -27.34 & 36.09 & 45.28 & 127.14 & 1 \\
PN\_8\_4\_1 & 00 38 46.17 & 41 28 19.4 & -243.4 & 6.0 & -15.40 & 43.67 & 46.31 & 109.42 & 1 \\
PN\_12\_4\_1 & 00 38 48.37 & 40 45 31.7 & -437.0 & 6.0 & -50.50 & 19.19 & 54.02 & 159.20 & 1 \\
PN\_12\_4\_2 & 00 38 49.51 & 40 44 25.4 & -573.2 & 11.6 & -51.29 & 18.38 & 54.48 & 160.28 & 1 \\
PN\_12\_1\_1 & 00 38 51.09 & 40 42 \ 3.8 & 84.6 & 13.5 & -53.07 & 16.80 & 55.66 & 162.43 & 1 \\
PN\_12\_4\_5 & 00 38 54.40 & 40 53 38.5 & -111.8 & 8.2 & -43.16 & 22.82 & 48.82 & 152.13 & 1 \\
\end{longtable}

\bsp

\label{lastpage}

\clearpage

\begin{thebibliography}{}
\bibitem[\protect\citeauthoryear{Arnaboldi et al.}{1996}]{arn96}
Arnaboldi, M., M\'{e}ndez, R. H., Capaccioli, M., Ciardullo, R., Ford, H. C., 
Gerhard, O., Hui, X., Jacoby, G. H., Kudritzki, R. P., Quinn, P. J. (1996).
\newblock {\em ApJ}, 472, 145.

\bibitem[\protect\citeauthoryear{Arnaboldi et al.}{1998}]{arn98} 
Arnaboldi, M., Freeman, K.C., Gerhard, O., Matthias, M., Kudritzki, R. P., 
M\'{e}ndez, R.H., Capaccioli, M., Ford, H. C. (1998).
\newblock {\em ApJ}, 507, 759.

%\bibitem[\protect\citeauthoryear{Barmby et al}{2001}]{bar01}
%Barmby, P., Huchra, J.P., Brodie, J.P. (2001)
%\newblock {\em AJ}, 121, 1482.

\bibitem[\protect\citeauthoryear{Belokurov et al}{2005}]{bel05}
Belokurov, V., An, J., Evans, N.W., Hewett, P., Baillon, P., Novati, S. Calchi, 
Carr, B.J., Cr\'{e}z\'{e}, M., Giraud-H\'{e}raud, Y., Gould, A., Jetzer, Ph., 
Kaplan, J., Kerins, E., Paulin-Henriksson, S., Smartt, S. J., Stalin, C.S., 
Tsapras, Y., Weston, M.J. (2005)
\newblock {\em MNRAS}, 357, 17.

\bibitem[\protect\citeauthoryear{Bender et al.}{2005}]{ben05}
Bender, R., Kormendy, J., Bower, G., Green. R., Thomas, J., Danks, A.C.,
Gull, T., Hutchings, J.B., Joseph, C.L., Kaiser, M.E., Lauer, T.R.,
Nelson, C.H., Richstone, D., Weistrop, D., Woodgate, B. (2005).
\newblock {\em ApJ}, 631, 280.

\bibitem[\protect\citeauthoryear{Braun}{1991}]{braun91}
Braun, R. (1991).
\newblock {\em ApJ}, 372, 54.

\bibitem[\protect\citeauthoryear{Carter and Jenkins}(1993)]{CJ93}
Carter, D., Jenkins, C.R., (1993).
\newblock {\em MNRAS}, 263, 1049.

\bibitem[\protect\citeauthoryear{Ciardullo et al.}{1989}]{cia89} 
Ciardullo, R., Jacoby, G. H., Ford, H. C., Neill, J. D. (1989).
\newblock {\em ApJ}, 339, 53.

\bibitem[\protect\citeauthoryear{Crane et al}{1992}]{cra92}
Crane, P.C., Dickel, J.R., Cowan, J. (1992)
\newblock {\em ApJ Lett}, 390, L9.

\bibitem[\protect\citeauthoryear{Crotts}{1992}]{cro92}
Crotts, A.P.S. (1992)
\newblock {\em ApJ Lett}, 399, 43.

\bibitem[\protect\citeauthoryear{Cseresnjes et al}{2005}]{cse05}
Cseresnjes, P., Crotts, A.P.S., de Jong, J.T.A., Bergier, A., Baltz, E.A., 
Gyuk, G., Kuijken, K., Widrow, L.M. (2005)
\newblock {\em ApJ Lett}, 633, 105.

\bibitem[\protect\citeauthoryear{de Jong}{2004}]{deJ04}
de Jong, J.T.A., Kuijken, K., Crotts, A.P.S., Sackett, P.D., Sutherland, W.J., 
Uglesich, R.R., Baltz, E.A., Cseresnjes, P., Gyuk, G., Widrow, L. M. (2004)
\newblock {\em A\&A}, 417, 461.

\bibitem[\protect\citeauthoryear{deVaucouleurs}{1948}]{dev48}
de Vaucouleurs, G. (1948)
\newblock {\em Ann Ap}, 11, 247.

\bibitem[de Vaucouleurs et al.(1991)]{dev91} de Vaucouleurs, G., de
Vaucouleurs, A., Corwin, H.~G., Buta, R.~J., Paturel, G., Fouque, P.,
(1991) 
\newblock Third Reference Catalogue of Bright Galaxies, Volume 1-3, XII,
Springer-Verlag Berlin Heidelberg New York.

\bibitem[\protect\citeauthoryear{Douglas}{2000}]{dou00}
Douglas, N.G., Gerssen, J., Kuijken, K., Merrifield, M.R. (2000)
\newblock {\em MNRAS}, 316, 795.

\bibitem[\protect\citeauthoryear{Douglas}{2002}]{dou02}
Douglas, N.G., Arnaboldi, M., Freeman, K.C., Kuijken, K., Merrifield, M. R., 
Romanowsky, A.J., Taylor, K., Capaccioli, M., Axelrod, T., Gilmozzi, R., 
Hart, J., Bloxham, G., Jones, D. (2002)
\newblock {\em PASP}, 114, 1234.

\bibitem[\protect\citeauthoryear{Dressler et al}{1999}]{dre99}
Dressler, A., Smail, I., Poggianti, B.M., Butcher, H.R.,
Couch, W.J., Ellis, R.S., Oemler, A. Jr. (1999)
\newblock {\em ApJ supp}, 122, 51.

\bibitem[\protect\citeauthoryear{Evans and Wilkinson}{2000}]{eva00}
Evans, N.W., Wilkinson, M.I. (2000).
\newblock {\em MNRAS}, 316, 929.

\bibitem[\protect\citeauthoryear{Evans et al}{2000}]{eva00b}
Evans, N.W., Wilkinson, M.I., Guhathakurta, P., Grebel, E.K., 
Vogt, S.S. (2000)
\newblock{\em ApJ} 540, 9.

\bibitem[\protect\citeauthoryear{Evans et al}{2003}]{eva03}
Evans, N. W., Wilkinson, M. I., Perrett, K. M., Bridges, T. J.
(2003)
\newblock {\bf ApJ}, 583, 752.

\bibitem[\protect\citeauthoryear{Federici et al}{1993}]{fed93}
Federici, L., Bonoli, F., Ciotti, L., Fusi-Pecci, F., Marano, B.,
Lipovetsky, V. A., Niezvestny, S. I., Spassova, N. (1993)
\newblock {\em A\&A}, 274, 87.

\bibitem[\protect\citeauthoryear{Ferguson et al}{2002}]{fer02}
Ferguson, A.M.N., Irwin, M.J., Ibata, R.A., Lewis, G.F., Tanvir, N.R. (2002)
\newblock {\em AJ}, 124, 1452.

\bibitem[\protect\citeauthoryear{Ford and Jakoby}{1978}]{fj78}
Ford, H.C., Jacoby, G.H. (1978)
\newblock {\em ApJ suppl}, 38, 351.

\bibitem[\protect\citeauthoryear{Gebhardt et al}{1996}]{geb96}
Gebhardt, K., Richstone, D.O., Ajhar, E.A., Lauer, T.R., Byun, Y.-I., 
Kormendy, J., Dressler, A., Faber, S.M., Grillmair, C., Tremaine, S. (1996)
\newblock {\em AJ}, 112, 105.

\bibitem[\protect\citeauthoryear{Gerssen et al}{2000}]{ger00}
Gerssen, J., Kuijken, K., Merrifield, M.R. (2000)
\newblock {\em MNRAS}, 317, 545.

\bibitem[\protect\citeauthoryear{Huchra et al}{1991}]{huchra1991}
Huchra, J.P., Brodie, J.P., Kent, S.M. (1991)
\newblock {\em ApJ}, 370, 495.

\bibitem[\protect\citeauthoryear{Hui et al}{1995}]{hui95}
Hui, X., Ford, H.C., Freeman, K.C., Dopita, M.A. (1995)
\newblock {\em ApJ}, 449, 592.

\bibitem[{Hurley-Keller et al}{2004}]{HK04}
Hurley-Keller, D., Morrison, H.L., Harding, P., Jacoby, G.H. (2004)
\newblock {\em ApJ}, 616, 804.

\bibitem[{Ibata et al.}{2001}]{ibata2001} Ibata, R., Irwin, M., 
Lewis, G., Ferguson, A.~M.~N., Tanvir, N. (2001) 
\newblock {\em Nature}, 412, 49. 

\bibitem[{Ibata et al.}{2004}]{ibata2004} Ibata, R., Chapman, S.,
Ferguson, A.M.N., Irwin, M.J.,Lewis, G.F., McConnachie, A. (2004)
\newblock {\em MNRAS}, 351, 117. 

\bibitem[{Ibata et al.}{2005}]{ibata2005} Ibata, R., Chapman, S.,
Ferguson, A.M.N., Lewis, G.F., Irwin, M.J., Tanvir, N.R. (2005) 
\newblock {\em ApJ}, 634, 287.

\bibitem[\protect\citeauthoryear{Irwin et al}{2005}]{irw05}
Irwin, M.J., Ferguson, A.M.N., Ibata, R.A., Lewis, G.F., Tanvir, N.R. (2005)
\newblock {\em ApJ Lett}, 628, L105.

\bibitem[\protect\citeauthoryear{Jablonka}{1998}]{jab98}
Jablonka, P., Bica, E., Bonatto, C., Bridges, T. J., Langlois, M., 
Carter, D. (1998)
\newblock {\em A\&A}, 335, 867.

\bibitem[\protect\citeauthoryear{Kent}{1989}]{kent1989}
Kent. S.M. (1989).
\newblock {\em AJ}, 97, 1614.

\bibitem[\protect\citeauthoryear{Kent et al}{1989}]{ken89b}
Kent, S.M., Huchra, J.P., Stauffer, J. (1989).
\newblock {\em AJ}, 98, 2080.

\bibitem[\protect\citeauthoryear{Kerins et al}{2001}]{ker01}
Kerins, E., Carr, B.J., Evans, N.W., Hewett, P., Lastennet, E.,
Le Du, Y., Melchior, A.-L., Smartt, S.J., Valls-Gabaud, D. (2001)
\newblock {\em MNRAS}, 323, 13.

\bibitem[\protect\citeauthoryear{Kerins et al.}{2003}]{ker03}
Kerins, E., An, J., Evans, N. W., Baillon, P., Carr, B. J., 
Giraud-H\'{e}raud, Y., Gould, A., Hewett, P., Kaplan, J., 
Paulin-Henriksson, S., Smartt, S.J., Tsapras, Y., Valls-Gabaud, D. (2003)
\newblock{\em ApJ}, 598, 993.

\bibitem[\protect\citeauthoryear{Kormendy}{1988}]{kor88}
Kormendy, J. (1988).
\newblock {\em ApJ}, 325, 128.

\bibitem[\protect\citeauthoryear{Kormendy and Bender}{1999}]{kb89}
Kormendy, J., Bender, R. (1999).
\newblock {\em ApJ}, 522, 722.

\bibitem[\protect\citeauthoryear{Lauer et al}{1993}]{lau93}
Lauer, T.R., Faber, S. M., Groth, E. J., Shaya, E. J., Campbell, B.,
Code, A., Currie, D. G., Baum, W. A., Ewald, S. P., Hester, J. J.,
Holtzman, J. A., Kristian, J., Light, R. M., Ligynds, C. R., 
O'Neil, E. J., Jr., Westphal, J. A. (1993)
\newblock {\em AJ}, 106, 1436.

\bibitem[\protect\citeauthoryear{McElroy}{1983}]{mcelroy}
McElroy, D.B. (1983).
\newblock {\em ApJ}, 270, 485.

\bibitem[\protect\citeauthoryear{Mendez et al}{2001}]{men01}
M\'{e}ndez, R.H., Riffeser, A., Kudritzki, R.-P., Matthias, M., 
Freeman, K.C., Arnaboldi, M., Capaccioli, M., Gerhard, O.E. (2001)
\newblock {\em ApJ}, 563, 135.

\bibitem[\protect\citeauthoryear{Merrett at al}{2003}]{mer03}
Merrett, H.R., Kuijken, K., Merrifield, M.R., Romanowsky, A.J., 
Douglas, N.G., Napolitano, N.R., Arnaboldi, M., Capaccioli, M., 
Freeman, K.C., Gerhard, O., Evans, N. W., Wilkinson, M. I., 
Halliday, C., Bridges, T.J., Carter, D. (2003)
\newblock {\em MNRAS}, 346, L62.

\bibitem[{Merrett et al.}{2003}]{Merrett2003} 
Merrett, H.~R., et al., (2003) 
\newblock {\em MNRAS}, 346, L62.

\bibitem[\protect\citeauthoryear{Merrett}{2006}]{mer06}
Merrett, H.R., {\sl et al.} (2006)
\newblock {\em MNRAS}, submitted.

\bibitem[\protect\citeauthoryear{Merritt}{1993}]{mer93a}
Merritt, D. (1993)
\newblock {\em ApJ}, 413, 79.

\bibitem[\protect\citeauthoryear{Merritt}{1996}]{mer96}
Merritt, D. (1996)
\newblock {\em AJ}, 112, 1085.

\bibitem[\protect\citeauthoryear{Merritt and Saha}{1993}]{mer93b}
Merritt, D., Saha, P. (1993).
\newblock {\em ApJ}, 409, 75.

\bibitem[\protect\citeauthoryear{Monet et al.}{1996}]{mon96}
Monet, D., Bird, A., Canzian, B., Dahn, C., Guetter, H.,
Harris, H., Henden, A., Levine, S., Luginbuhl, C., Monet, A. K. B.,
Rhodes, A., Riepe, B., Sell, S., Stone, R., Vrba, F., Walker, R. (1996).
\newblock {\em A Catalog of Astrometric Standards}, 
U.S. Naval Observatory, Washington DC.

\bibitem[\protect\citeauthoryear{Moore et al}{2001}]{moo01}
Moore, B., Calc\'{a}neo-Rold\'{a}n, C., Stadel, J., Quinn, T.,
Lake, G., Ghigna, S., Governato, F. (2001)
\newblock {\em Phys Rev D}, 64, 3508.

\bibitem[{Morrison et al.}{2004}]{Morrison2004} Morrison, H.~L.,
Harding, P., Perrett, K., Hurley-Keller, D., (2004) 
\newblock {\em ApJ}, 603, 87.

\bibitem[\protect\citeauthoryear{Ostriker et al.}{1974}]{ost74}
Ostriker, J.P., Peebles, P.J.E., Yahil, A. (1974)
\newblock {\em ApJ Lett}, 193, L1.

\bibitem[\protect\citeauthoryear{Parry}{1994}]{par94}
Parry, I.R., Lewis, I.J., Sharples, R.M., Dodsworth, G.N., 
Webster, J., Gellatly, D.W.; Jones, L.R., Watson, F.G. (1994)
\newblock {\em Proc SPIE}, 2198, 125.

\bibitem[\protect\citeauthoryear{Peng et al}{2004}]{peng04}
Peng, E.W., Ford, H.C., Freeman, K.C. (2004)
\newblock {\em ApJ}, 602, 685. 

\bibitem[\protect\citeauthoryear{Perrett et al}{2002}]{per02}
Perrett, K.M., Bridges, T.J., Hanes, D.A., Irwin, M.J., Brodie, J.P.,
Carter, D., Huchra, J.P., Watson, F. G. (2002)
\newblock {\em AJ}, 123, 2490.

\bibitem[\protect\citeauthoryear{Perrett et al}{2003}]{per03}
Perrett, K.M., Stiff, D.A., Hanes, D.A., Bridges, T.J. (2003)
\newblock {\em ApJ}, 589, 790.

\bibitem[\protect\citeauthoryear{Poggianti et al}{2006}]{pog06}
Poggianti, B.M., von der Linden, A., De Lucia, G., Desai, V.,
Simard, L., Halliday, C., Aragon-Salamanca, A., Bower, R., Varela, J.,
Best, P., Clowe, D.I., Dalcanton, J., Jablonka, P., Milvang-Jensen, B.,
Pello, R., Rudnick, G., Saglia, R., White, S.D.M., Zaritsky D. (2006)
\newblock {\em ApJ} in press, (astro-ph/0512391).

\bibitem[\protect\citeauthoryear{Pritchet and van den Bergh}{1994}]{Pri94}
Pritchet, C.J., van den Bergh, S. (1994)
\newblock {\em AJ}, 107, 1730.

\bibitem[\protect\citeauthoryear{Riffeser et al}{2003}]{rif03}
Riffeser, A., Fliri, J\"{u}rgen, Bender, R., Seitz, S., G\"{o}ssl, C.A.
(2003)
\newblock {\em ApJ}, 599, 17.

\bibitem[\protect\citeauthoryear{Roberts and Whitehurst}{1975}]{rob75}
Roberts, M.S., Whitehurst, R.N. (1975)
\newblock {\em ApJ}, 201, 327.

\bibitem[\protect\citeauthoryear{Rubin and Ford}{1970}]{rub70}
Rubin, V.C., Ford, W.K. Jr. (1970)
\newblock {\em ApJ}, 159, 379.

\bibitem[\protect\citeauthoryear{van der Marel et al.}{1994}]{vdm94}
van der Marel, R.P., Rix, H.-W., Carter, D., Franx, M., White, S.D.M.,
de Zeeuw, T. (1994)
\newblock {\em MNRAS}, 268, 521.

\bibitem[\protect\citeauthoryear{Widrow and Dubinski}{2005}]{wid05}
Widrow, L.M., Dubinski, J. (2005)
\newblock {\em ApJ}, 631, 838.

\bibitem[\protect\citeauthoryear{Worswick et al}{1995}]{wor95}
Worswick, S.P., Gellatly, D.W., Ferneyhough, N.K., King, D.L., Weise, A.J., 
Bingham, R.G., Oates, A.P. (1995)
\newblock {\em Proc SPIE}, 2476, 46.

\end{thebibliography}
\end{document}